\documentclass[final,3p,times]{elsarticle}
\usepackage{lineno,hyperref}
\usepackage{calc}
\usepackage{epsfig,amsmath,amssymb,graphics,graphicx}

\usepackage{amscd}
\usepackage{hyperref}
\usepackage{amsfonts}
\usepackage{epsfig}

\modulolinenumbers[5]

\journal{Annals of Physics}

\begin{document}

\begin{frontmatter}

\title{The new model method of the electrostatic screening describing: three-component system of the "closed" type}

\author{A. A. Mihajlov, Lj. M. Ignjatovi{\' c} and V. A. Sre{\' c}kovi{\' c}}
\address{Institute of Physics, Belgrade University, Pregrevica 118,
Zemun, 11080 Belgrade, Serbia}

%
%

\begin{abstract}
Within the problem of the finding of the  mean potential energy of the charged particle in the plasma
in this work a classification of physical systems (electrolytes, dusty plasmas, plasmas)
is made based on consideration, or lack thereof, of a few special additional conditions. The system
considered here, as well as other systems which are described with those additional conditions imposed
are treated as the systems of the "closed" type, while the systems where
those conditions are fully neglected - as being of the "open" type.
In the our previous investigation one- and two component systems were examined. Here,
as the object of investigation
fully ionized electron-ion plasma with positive charged ions of two different kinds, including
the plasmas of higher non-ideality, is chosen for the first time.
For such plasma a new self-consistent model method of describing electrostatic screening is developed
which includes all the necessary additional conditions.
By this method such extremely significant phenomena as electron-ion and ion-ion correlations are included in the used model.
The characteristics of the considered plasmas in a wide region of electron densities and temperatures are calculated.
All the obtained results, including a comparison with systems of the "open" type are specifically discussed.

\end{abstract}


\end{frontmatter}


\section{Introduction}
\label{sec:intro}

Thematically, this work is the natural extension of the research of the mechanism of plasma's inner
electrostatic screening whose results are presented in the papers \cite{mih08,mih09a, mih09b},
named here Part 1, Part 2 and Part 3, respectively.
Namely, in these papers the single- and two- component systems were discussed,
and here we will consider for the first time systems of the next level of complexity i.e.
three-component systems which contain free electrons and positive charged
ions of two different kinds. However, as for the method of investigation, we shall
emphasize straight away that it will be completely different than in the previous papers.
Because of that, we will remind that the conducted research had the following task:
to investigate, within the problem of the finding of the  mean potential energy of the charged particle in the plasma,
whether the physical model of plasma's inner electrostatic screening,
introduced in \cite{deb23}, is already exhausted by Debye-H\"{u}ckel's (DH) method,
described in the same paper, or it still allows for development of an alternative one.
As in the previous papers here we keep in mind the electrostatic screening in fully ionized
plasmas.

Although the paper \cite{deb23} was devoted to electrolytes it left profound trace in plasma physics and in
adjacent disciplines (\cite{ich73,kit77,dra65,kra86}, see also \cite{dim89}). Its influence is felt
even today in various fields of plasma physics (laboratory,  astrophysical, ionosphere).
So, in numerous papers direct DH  or DH-like methods are used, as well as such products of theirs
as DH potential and DH radius (see Inv.Bre. refs. and \cite{shu08,zha04,lin10,lin11}).
All this sparked the interest for the possibility of going beyond the sphere of influence
of \cite{deb23} and development of the mentioned alternative method. Also,
another stimulus existed for development of alternative methods, connected with finding
a characteristic length grater than Debye radius (\cite{gun84,vit01,gun72,gun76,gol78,gun83}, see also Part 3).

Let us remind that the essential properties of the mentioned model are the following:

\textbf{-} the presence of an immobile probe particle, which represents one
kind of charged particles in the real system (plasma, electrolyte);

\textbf{-} treatment of the considered components which contain free charged
particles of different kinds as ideal gases in states of
thermodynamical equilibrium, without the assumption about equality of
all temperatures;

\textbf{-} treatment of the existing total electrostatic field in the considered
system as external with respect to the considered ideal gas;

\textbf{-} finally, among the properties of this model is usage, as its relevant mathematical apparatus,
of equations which describe the mean local electrostatic field
and the conditions of conservation of thermodynamical equilibrium for the considered components.
As in the previous papers, this model is treated here as the basic one.

Let us remind that in the source work \cite{deb23} the authors did not assume the electro-neutrality
of the considered system, it was obtained as an organic property of the DH solution. This fact deserves
special attention, since the mentioned electro-neutrality
was automatically provided regardless of the fact that DH solution is a highly approximated one
and has a few nonconditional negative properties (see discussion in Part 2).
It seems that in \cite{deb23} the electro neutrality was obtained as a present, which
was justified by the ingeniosity of the described basic model. In connection with this
it is necessary to keep in mind that any improvement of DH method means
absence of an automatical electro-neutrality of the considered system and
requires completion of the basic model by an additional condition of electro-neutrality
of the system. Consequently, this fact is taken into account in this investigation from the beginning.

The task formulated above itself enforced a special role of the mentioned DH method
as a referent one whose predictions shall be compared with a possible new method
that would arise as the result of the undertaken investigation.
In accordance with this, the main aim of our previous research became
creation of a "self-consistent" method of describing the mentioned electrostatic screening mechanism
which is completely free of DH method's disadvantages.
Let us note that the definition of a "self-consistent" method implies that all the relevant characteristic are
determined within this method itself and expressed only through its
basic parameters i.e. the particle densities, temperature, etc.

However, it has been shown that, except for the case of single-component system
(e.g. electron gas on a positively charged background),
it was very difficult to finish the whole procedure of eliminating the disadvantages
of DH method in a "self-consistent" way.
Analysis which was done later convinced us that this result is not accidental,
since the outer differences between the DH and the presented method were not practically significant, and
the principal differences between these methods are of conceptual nature.

It was connected with application of additional conditions, equivalent to the conditions
of conservation of particle numbers in finite systems, which are applicable in the case of the basic model (see Section 3).
Namely, it can be shown that the only important fact here is that in the presented method
relevant additional conditions are taken into account, while within DH method they are completely neglected.
Because of the above mentioned, the DH method lost its especial significance for this research.
In accordance with this, further on we will treat it
only as an example of a method where the mentioned additional condition are neglected and we will not deal with its disadvantages.
Consequently, apart of the finding of the mean potential energy of the charged particles in the plasmas, the direct objective of this research became development of a self-consistent method of describing the electrostatic screening in the considered three-component system where the relevant additional conditions are included from the beginning.
Let us note that in such form our task has practically shifted from a purely physical to
a mosly mathematical domain, as physics is now due to appear only at the point of presentation of
the final results. We have certainly taken good care in order that the approximations used within the
developed procedures be physically sound, so that application of the obtained results to
real plasmas could make sense.

\begin{figure*}
\begin{center}
\includegraphics[width=0.48\textwidth]{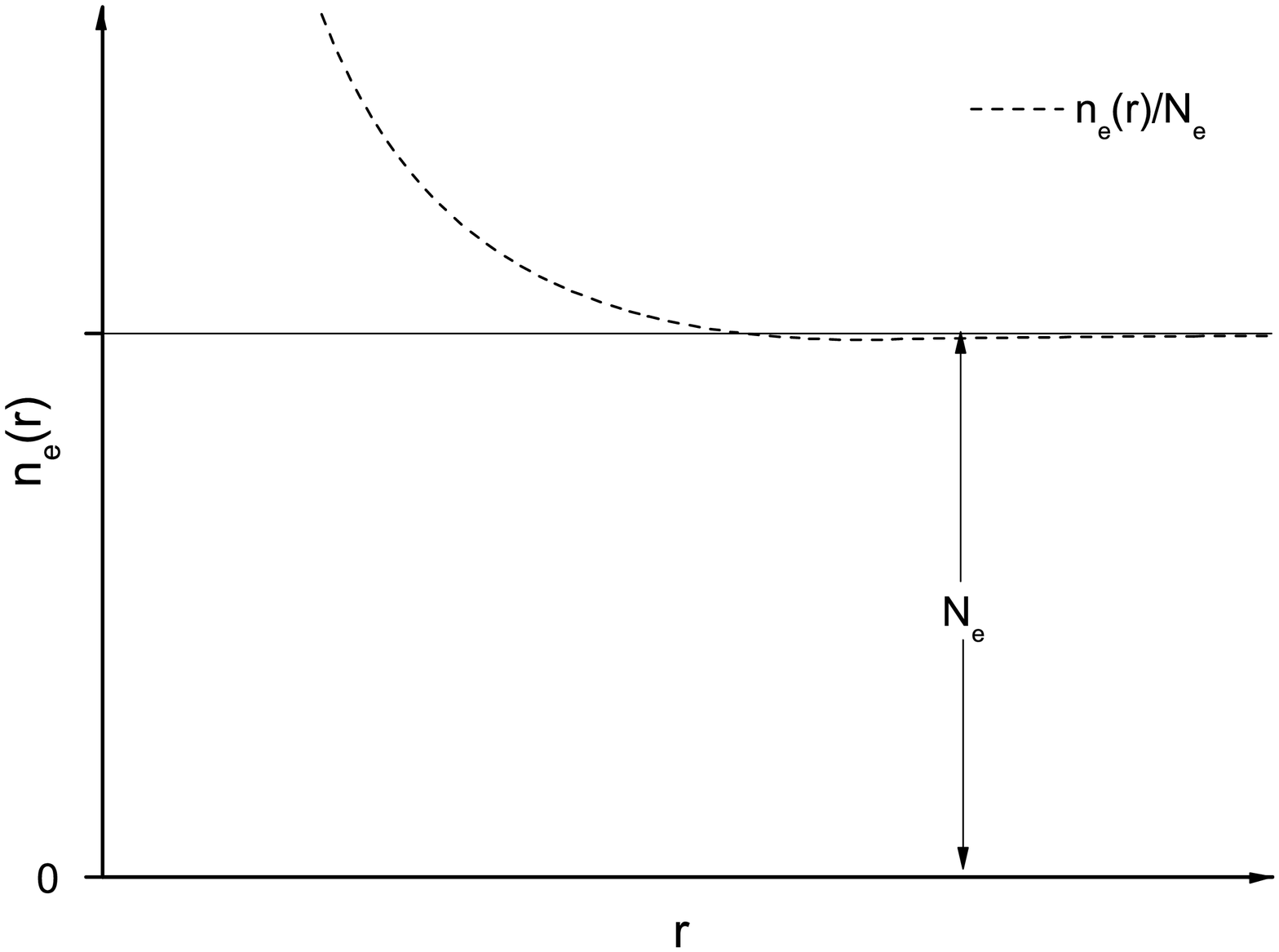}
\includegraphics[width=0.48\textwidth]{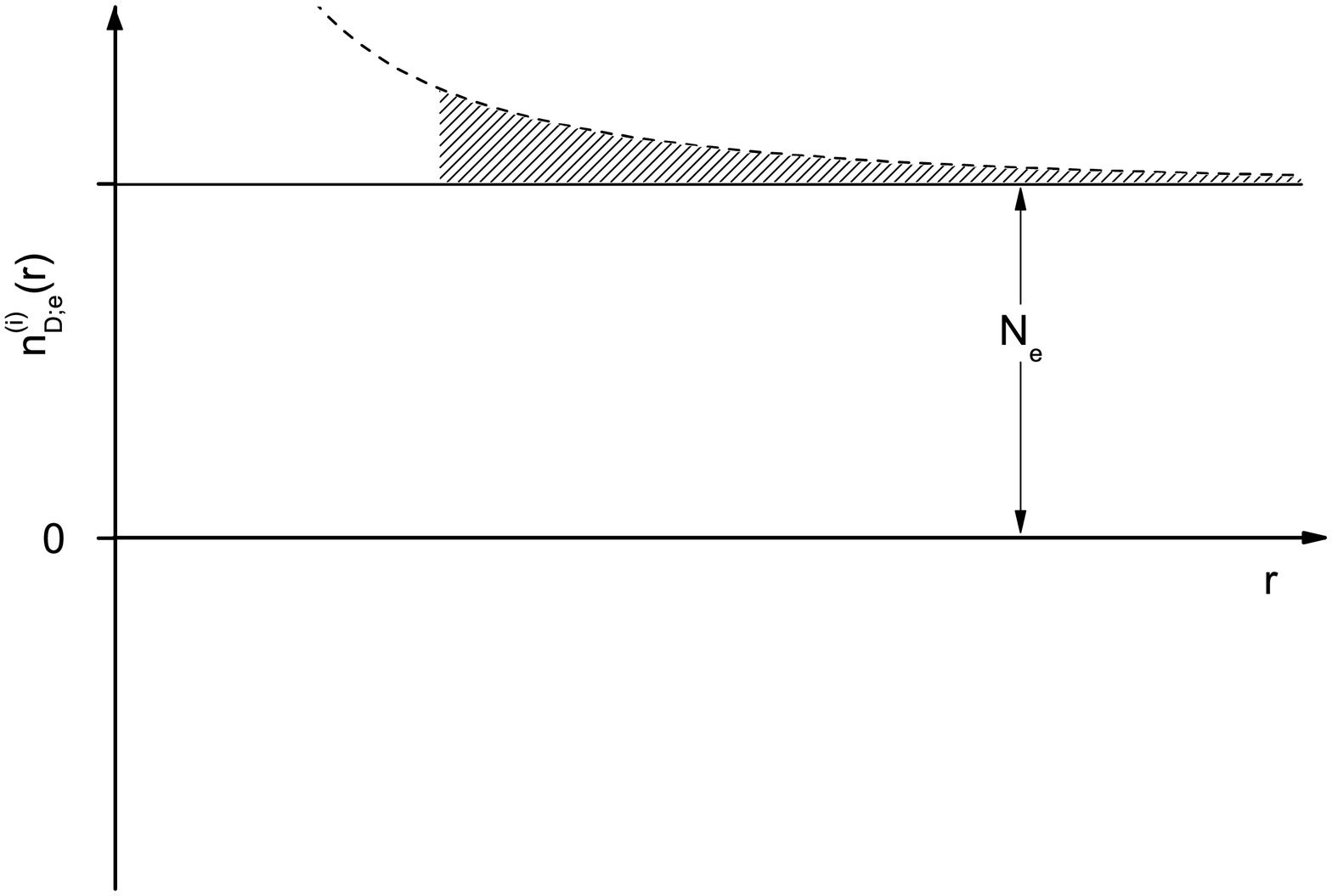}
\caption{\hspace{0.3in} a) \hspace{2in} b) \newline a) The behavior of the electron in the case of the electron ion plasma with the probe particle whose charge is equal to ions
which is obtain by means the needed additional conditions, b) The behavior of the DH electron in the case of the electron proton plasma with the probe particle whose charge is equal to protons.}
\label{fig:htbpprof}
\end{center}
\end{figure*}
In order to show the differences between the results of applying and
neglecting the relevant additional conditions, we will remind of the behaviour
of electron density in two cases of the electron-ion plasmas with the
probe particle whose charge is equal to that of an ion, which was considered in Part 2,
and is illustrated here by Figs.~\ref{fig:htbpprof}a and \ref{fig:htbpprof}b. First of them illustrates
application of the method developed in Part 2, which makes sure that the area of
higher electron density is followed by an area of its lowering. This is in
accordance with the role of the probe-particle approximation: the situation in the vicinity of the probe
particle could be considered as reflection of the real situation in the vicinity
of any ion in the considered plasma, and should enable usage of the results illustrated
by the figure to the case of the real plasma, since they don't influence
the mean electron density. The systems with similar behaviour of the electron density
are treated here as the "closed" ones. In the figure 1b the behaviour
of the DH electron density is shown in electron-proton plasma with the probe particle
whose charge is equal to proton's. From this figure monotonous increasing of the electron density can be seen with
decrease of the distance from the probe particle from infinity to zero. In the considered case such behaviour
causes creation of an excess of 1/2 electron and 1/2 proton in the vicinity of the probe particle.
This phenomenon, which is unacceptable from the point of view of a method whith includes additional conditions,
will be discussed especially in the Section 7.

The systems with a similar behaviour of the electron density are treated here as "open" ones.
Let us note that only the systems of the "closed" type are considered here in detail.
The difference between "open" and "closed" systems will be discussed again at the end of this paper.

This work is dedicated to plasmas which are treated as fully ionized,
including the plasmas of higher non-ideality. The
region of electron densities from $10^{16}$cm$^{-3}$ to $10^{20}$cm$^{-3}$ is considered, and of temperatures from
$1\cdot 10^{4}$ K to $3\cdot 10^{4}$ K.
The material presented in this paper is distributed in seven Sections
and five sections in Supplementary material. Section~\ref{sec:tas} contains the basic
assumptions of the theory, Section 3, Section 4 and Section 5 - the
description of the methods developed for the considered three-component systems in
the case of positively and negatively charged probe particles respectively,
Section~\ref{sec:par} - the procedure used for determination of the theory
parameters, Section~\ref{sec:pso} - the results and the necessary discussion,
Supplementary material A, B, C contain a description of the
procedures of obtaining some auxiliary expressions whose inclusion in the
main part of the text could overload it, Supplementary material D - online materials with the tables
and Supplementary material E - the future perspective of DH method.

\section{Theory assumptions}
\label{sec:tas}

\subsection{The initial system and basic characteristic}
\label{sec:tas}

A stationary homogeneous and isotropic system $S_{in}$ is taken here as the
initial model of some real physical objects, suitable for applications of the
results of this research. It is assumed that $S_{in}$ is constituted by a mix of
a gas of free electrons and two gases of free ions of different kinds
with positive charges $Z_{1}e$ and $Z_{2}e$, where $Z_{1,2}=1,2,3,...$, and
$e$ is the modulus of the electron charge. The electron charge, $-e$, will
be denoted also by $Z_{e}e$, where $Z_{e}= -1$. We will consider that these gases
are in equilibrium states with: the mean densities $N_{1}$ and $N_{2}$
and the temperatures $T_{1}=T_{2}=T_{i}$ for the ions, and the mean density
$N_{e}$ and the temperature $T_{e}\ge T_{i}$ for the electrons. All particles are
treated as point-like, non-relativistic objects and their spins are taken into
account only as factors which influence the chemical potentials of the
considered gases. Satisfying the the condition
\begin{equation}
\label{eq:1} Z_{1}e \cdot N_{1} + Z_{2}e \cdot N_{2} - e \cdot
N_{e} = 0,
\end{equation}
is understood,
which provides local quasi-neutrality of the system $S_{in}$. Let us
emphasize that the case $Z_{1}=Z_{2}$ is also considered here (see subsection
\ref{sec:pso1}, Section \ref{sec:pso}), since it reflects the existence of some
real systems with two physically different kinds of the ions with the same
charge, e.g. H$^{+}$ and D$^{+}$, or H$^{+}$ and He$^{+}$ etc.

In the subsequent considerations several characteristic lengths are used
which are connected with the parameters $N_{1,2,e}$ and $Z_{1,2}$,
namely the Wigner-Seit's (WS) radii and the so-called ion self-spheres
radii (see Part 2), denoted here by $r_{1,2,e}$ and $r_{s;1,2}$
respectively. They are defined by the relations
\begin{equation}
\label{eq:1a} \frac{4\pi}{3}\cdot r_{1,2,e}^{3} =\frac{1}{N_{1,2,e}}, \quad
\frac{4\pi}{3}\cdot r_{s;1,2}^{3} =\frac{Z_{1,2}}{N_{e}}, \quad
r_{s;e}\equiv r_{e}.
\end{equation}
From here it follows that Eq. (\ref{eq:1}) can be presented in the form
\begin{equation}
\label{eq:1b} p_{1} + p_{2} = 1, \qquad p_{1,2}\equiv\frac{N_{1,2}\cdot
Z_{1,2}}{N_{e}}= N_{1,2}\cdot \frac{4\pi}{3}r_{s;1,2}^{3},
\end{equation}
where the parameters $p_{1}$ and $p_{2}$ describe the primary distribution of the
space between the self-spheres of all ions of the first kind and all ions of the second kind.

\subsection{The auxiliary systems}
\label{sec:acces}

{\bf The properties and system conditions.} In accordance with the basic
model and the composition of the system $S_{in}$, the electrostatic
screening of the charged particles is modeled here
in three corresponding auxiliary systems. It is assumed that each of
them contains: the electron component, two ion components with the
same charges $Z_{1}e$ and $Z_{2}e$, and one immobile probe particle
with the charge $Z_{p}e$, which is fixed in the origin of the used
reference frame (the point $O$).

As in Part 1 and 2 only such cases are studied here where the probe
particle can represent one of the charge particles of the system
$S_{in}$, e.g. when $Z_{p} = Z_{1}$, $Z_{2}$ and $Z_{e}$. Two
ion cases are denoted below with $(i1)$ and $(i2)$,
the electron case - with $(e)$, while the corresponding auxiliary
systems are denoted with $S_{a}^{(1)}$, $S_{a}^{(2)}$ and $S_{a}^{(e)}$
respectively.

All systems $S_{a}^{(1,2,e)}$ are treated below as isotropic and
characterized by the corresponding mean local ion and electron densities:
$n_{1}^{(1,2,e)}(r)$, $n_{2}^{(1,2,e)}(r)$ and $n_{e}^{(1,2,e)}(r)$, which retain
the properties of the corresponding components in the system $S_{in}$
and satisfy the boundary conditions
\begin{equation}
\label{eq:3}  \lim\limits_{r \to \infty } n_{1}^{(1,2,e)}(r) = N_{1},
\quad \lim\limits_{r \to \infty } n_{2}^{(1,2,e)}(r) = N_{2}, \quad
\lim\limits_{r \to \infty } n_{e}^{(1,2,e)}(r) = N_{e},
\end{equation}
where $r = \vec{r}$, and $\vec{r}$ is the radius-vector of the observed point. Their other
necessary characteristics are the mean local charge density $\rho^{(1,2,e)}(r)$
defined by relation
\begin{equation}\label{eq:2}
\qquad \rho^{(1,2,e)}(r) = Z_{1}e\cdot n_{1}^{(1,2,e)}(r) +
Z_{2}e\cdot n_{2}^{(1,2,e)} - e\cdot n_{e}^{(1,2,e)}(r),
\end{equation}
and the mean local electrostatic potential $\Phi^{(1,2,e)}(r)$,
which is treated as the potential of the external electrostatic
field.
Then, we will take into account the fact that $\Phi ^{(1,2,e)}(r)$ and
$\rho^{(1,2,e)}(r)$ have to satisfy the Poisson's equation
\begin{equation}
\label{eq5}  \nabla^{2} \Phi ^{(1,2,e)} = - 4\pi \left[ Z_{1,2,e}e\cdot
\delta(\vec r) + \rho ^{(1,2,e)}(r) \right],
\end{equation}
where $\delta(\vec r)$ is the three-dimensional delta function
\cite{ich73}, and $0 \le r \le \infty$. Satisfying the boundary conditions
\begin{equation}\label{eq:5}
\lim\limits_{r \to \infty} \Phi^{(1,2,e)}(r) = 0, \quad \left
|{\varphi^{(1,2,e)}} \right | < \infty; \smallskip \varphi^{(1,2,e)}
\equiv \lim\limits_{r \to 0} [\Phi^{(1,2,e)}(r) - \frac{Z_{1,2,e}e}{r}],
\end{equation}
is assumed,
which guaranties physical sense of the mentioned electrostatic potential and connection
with the system $S_{in}$, and is compatible with the electro-neutrality condition of the auxiliary systems.
Since $\varphi ^{(1,2,e)}$ is the mean electrostatic potential in the
point $O$, the quantity
\begin{equation}
\label{eq8}  U^{(1,2,e)} = Z_{1,2,e} e \cdot \varphi ^{(1,2,e)}.
\end{equation}
is the mean potential energy of the probe particle is just the searched mean potential
energy of the probe particle. In a
usual way $U^{(1,2)}$ and $U^{(e)}$ are treated as approximations to the
mean potential energies of the ion and electron in the initial system
$S_{in}$.

In accordance with the basic model the electron and all ion components
of all auxiliary systems are treated as ideal gases.
Therefore we will encompass the characteristics of the auxiliary systems by
chemical potentials $\mu_{1,2}(n_{1,2}(r),T_{i})$ and $\mu_{e}(n_{e}^{(e)}(r),T_{e})$
of the corresponding ideal ion and electron gases,
which can depend on the corresponding particle spins; and
by their boundary values, i.e. $\mu_{1,2}(N_{1,2}^{(1,2)},T_{i})=
\lim\limits_{r\to\infty}\mu_{1,2}(n_{1,2}(r),T_{i})$
$=$
$\lim\limits_{r\to\infty}\mu_{1,2}(n_{1,2}^{(2,1)}(r),T_{i})$
and $\mu_{e}(N_{e},T_{e})=
\lim\limits_{r\to\infty}\mu_{e}(n_{e}^{(e)}(r),T_{e})$.
However, in accordance with Part 2, the ion components in the
auxiliary system $S_{a}^{(e)}$ has to be considered in a special way which is described in the Section
"The case (e)".

\noindent {\bf The system equations.}
It can be shown that on the basis of the procedure which was developed and described in detail in Part 1 and 2
for the case of the system which is electro-neutral as a whole, it is possible to switch from the Poisson's
equation to the equation for the potential $\Phi^{(1,2,e)}(r)$
which is more suitable for further consideration.
This equation is given by
\begin{equation}\label{eq:6}
\Phi ^{(1,2,e)}(r)= -4\pi \int\limits_{r}^{\infty} {\rho^{(1,2,e)}(r)}
(r')\left({\frac{1}{r} - \frac{1}{r'}} \right) r'^{2}dr',
\end{equation}
and is taken here in such a form.

In order to find other necessary equations, we will consider the conditions of
conservation of thermodynamical equilibrium (conservation of the electro-chemical potential) for those components
which are represented by the corresponding probe particles, namely
\begin{equation}\label{eq:7}
\mu_{\delta}(n_{\delta}^{(\delta)}(r),T_{\delta})+Z_{\delta}e
\cdot \Phi^{(\delta)}(r)=
\mu_{\delta}(n_{\delta}^{(\delta)}(r_{st}),T_{\delta})+
Z_{\delta}e\cdot \Phi^{(\delta)}(r_{st}),
\quad \delta= 1, 2, e,
\end{equation}
where, in accordance with the basic model, $\Phi^{(1,2,e)}(r)$ is treated as the
potential of the external electrostatic field, and $r_{st}$ is the
distance from the point $O$ of the chosen fixed (starting) point: $0<r_{st}\le \infty$. From here, by means
the usual linearization procedure, the necessary equations for the particle
densities $n_{\delta}^{(\delta)}(r)$ are obtained in the form
\begin{equation}\label{eq:121}
n_{\delta}^{(\delta)}(r) - n_{\delta}^{(\delta)}(r_{st}) = -
\frac{Z_{\delta}e}{\partial \mu_{\delta}/\partial N_{\delta}}
\cdot\left[\Phi^{(\delta)}(r) - \Phi^{(\delta)}(r_{st}) \right], \quad
\frac{\partial\mu_{\delta}}{\partial N_{\delta}} \equiv
\left[\frac{\partial\mu_{\delta} (n,T_{\delta})}{\partial
n}\right]_{n = N_{\delta}},
\end{equation}
which is applicable under the condition
\begin{equation}
\label{eq13a}
\frac{|n_{\delta}^{(\delta)}(r)-N_{\delta}|}{N_{\delta}} \ll 1.
\end{equation}
Here we will use the fact that such equations can be applied not only to the
classical cases, but also to the quantum-mechanical ones, including
the case of ultra-degenerated electron gas \cite{kit77,kra86}.

Let us note that the corresponding equations for the ion
densities $n_{1,2}^{(e)}(r)$ are obtained and discussed in
Section~\ref{sec:mdme}, and for $n_{2,1}^{(1,2)}(r)$ - in \ref{sec:mdmi}.

\noindent {\bf The additional conditions.}
In Part 1 and 2 the conditions were already introduced for the component which is represent by the probe particle and
in which the charge of the probe particle appears.
In the considered three-component case these conditions are given by
\begin{equation}
\label{eq:13}
\int\limits_0^\infty {\left[N_{1,2,e} -
n_{1,2,e}^{(1,2,e)} (r) \right] \cdot 4\pi r^2} dr =\int\limits_0^\infty
{\left[1-\frac{n_{1,2,e}^{(1,2,e)}(x \cdot
r_{1,2,e})}{N_{1,2,e}} \right] \cdot 3  x^{2}}dx = 1,
\end{equation}
where $x\equiv r/r_{1,2,e}$. This means that
the ratio in the expression inside the square brackets is of the order of magnitude of 1.
This equation is especially important since it provides the continuity of the model.
In order to show this fact it is enough to consider the situation  when the charge
density $n_{1,2}^{(2,1)}$ is negligible and the considered three systems for physical reasons
can be treated as a two-component system. Then, as an approximation, we could replace the
electron component by the negatively charged nonstructural background
and come back to a one-component system. It is important that the single-component
case is by now used for mathematical modeling of the plasma internal electrostatic
screening (Iosilevskiy 2011, private communication).
Let us note that in the single-component case (which was not considered in \cite{deb23})
this first condition can be used instead of the electro-neutrality condition.

Then in Part 2 it the
additional conditions for the electron density were introduced. Here
the corresponding conditions are given by the equations
\begin{equation}
\label{eq:14}  \int\limits_0^\infty {\left[N_{1,2} - n_{1,2}^{(e)}(r)\right]
\cdot 4\pi r^2}dr=\int\limits_0^\infty {\left[N_{e} - n_{e}^{(1,2)}(r)\right]
\cdot 4\pi r^2}dr = 0.
\end{equation}
The additional conditions for the ion components $n_{1}^{(2)}(r)$ and $n_{2}^{(1)}(r)$
are taken into consideration analogously to the electron components,
since there are no principal differences between them. The corresponding
relation are given by
\begin{equation}\label{eq:three}
\int\limits_0^\infty {\left [N_{1} - n_{1}^{(2)}(r)\right]\cdot
4\pi r^2} dr = \int\limits_0^\infty {\left [N_{2}-n_{2}^{(1)}(r)\right]\cdot
4\pi r^2} dr = 0.
\end{equation}
As it was mentioned (Introduction), the presented additional conditions
are fully compatible with the basic model. Let us note that in this work the considered physical systems,
as well as other physical systems which are described by means of additional conditions (\ref{eq:13})-(\ref{eq:three}),
are treated as systems of the "closed" type, while the physical systems where they are fully neglected-
as the systems of the "open" type.
The significance of these conditions and consequences of their application are discussed in Section 7.

As it is known, in our investigation we have to take care
that our results be compared with the results of DH method, which provides
electro-neutrality of the considered system as a whole. In principle
this would justify introducing into consideration the corresponding
electro-neutrality condition, namely
\begin{equation}\label{eq:4}
Z_{1,2,e}e+\int\limits_0^\infty {\rho ^{(1,2,e)}(r)}\cdot 4\pi r^2dr = 0.
\end{equation}
However the fact is used that simultaneous satisfaction
of the conditions (\ref{eq:13}) and (\ref{eq:14}) and (\ref{eq:3}) automatically
provides satisfaction of this condition.
Therefore this condition will not be used within this work.

\section{(i1) and (i2) cases: auxiliary expressions }

\subsection{Needed remarks}

Practically from the beginning of the research
insufficiency (unsatisfactoriness) of equations
and system equations from the previous section was a serious problem.
Concerning the electron component, this problem was solved in Part 2 by elaboration of a complete acceptable procedure for its determination.
We will take advantage of this fact and use the already obtained results in the next section with minimal necessary discussion.

As for the ion component, a similar problem is solved in this paper, and  we
need to present all the relevant considerations here. Because of the abundance of the necessary material a part of it
is displayed in Supplementary material A and B.

For the sake of further consideration we will note that here, apart
from $r_{s;1,2}$, $r_{1,2}$ and $r_{e}$, other
characteristics lengths will be also used, namely: $l_{s;1,2}$ and $r_{b;1,2} > l_{s;1,2}$
in the cases $(i1)$ and $(i2)$, and $l_{e;1,2}$ in the
case $(e)$. So, we will treat the regions $0<r<l_{s;1,2}$ and
$r_{b;1,2}<r<\infty$ as those of small and large $r$ respectively
and, consequently, the regions $l_{s;1,2} < r <r_{b;1,2}$ - as those of middle $r$.

\subsection{The electron densities}

{\bf The region of large and middle $r$.} In
Part 2 it the model of the electron component in the
corresponding auxiliary system was introduced which can be applied in the case
of the considered $S_{a}^{(1,2)}$ systems.
It was obtained on the basis of the approximation where
the heavy particles (ions) were treated as as an ideal gas with respect to the fixed probe particle,
but as immobile with respect to the free light particles (electrons). Let us note that
similar approximations were very often used in the physics of plasma (see e.g. \cite{vit90}).
Apart from that, in the region of small $r$
(the vicinity of the positively charged probe particle), equality of
all positively charged ions of the same kind was used, as well as the symmetry of the considered system.
Typical behaviour of the electron density in the three-component case is presented in Fig.~\ref{fig:htbp11}.

Within this model the changes of the electron
density in the regions of large and middle $r$ were expressed through changes of the
total ion-charge density. For that purpose it was taken into account that in the considered regions
of $r$ the changes of electron and  ion densities are small and can be very well described
in the linear approximations.

One can see that the mentioned model can be automatically
applied to systems with more positively charged ion components.
Namely, it is enough to use in the just mentioned relation the changes
of the positive-charge densities of all the existing ion components.
So, in the considered three-component case, the relevant relation should have the form:
$[n_{e}^{(1,2)}(r)- N_{e}] \sim [(Z_{1} \cdot n_{1}^{(1,2)} +
Z_{2}\cdot n_{2}^{(1,2)}) - (Z_{1}\cdot N_{1} + Z_{2}\cdot N_{2})],$
whence it follows that the necessary expression for the electron densities
in the regions of middle and large $r$ can be taken in the form
\begin{equation}\label{eq:22}
n_{e}^{(1,2)}(r) = N_{e}\cdot (1-\alpha_{e;1,2}) + \alpha_{e;1,2} \cdot
[Z_{1} n_{1}^{(1,2)}(r) + Z_{2} n_{2}^{(1,2)}],
\end{equation}
where by $\alpha_{e;1}$ and $\alpha_{e;1}$ the corresponding proportionality coefficients
are denoted, which are treated here as the corresponding electron-ion correlation coefficients
($0 < \alpha_{e;1,2}<1$).\\

\noindent {\bf The region of small $r$.}
For such a region another adequate procedure for
determination of electron density was developed in Part 2, which is also applicable to
systems with more positively charged ion components. In the considered case
the parameters $l_{s;1,2}$ are used for that procedure instead of $r_{s;1,2}$.
It is understood that $l_{s;1,2}$ have to satisfy the condition
\begin{equation}
\label{eq:1c} N_{1}\cdot \frac{4\pi}{3}l_{s;1}^{3} +
N_{2}\cdot \frac{4\pi}{3}l_{s;2}^{3} = 1,
\end{equation}
which reflects the fact that transition from $r_{s;1,2}$ to
$l_{s;1,2}$ has to represent only redistribution of the space
between the ion components, which means the equality of the left
sides of Eqs. (\ref{eq:1b}) and (\ref{eq:1c}). Let us note that in the case
$Z_{1}=Z_{2}$ we have it that $l_{s;1,2}=r_{s;1,2}$.

Within the mentioned procedure the electron densities $n_{e}^{(1,2)}(r)$ in the
regions $0<r<l_{s;1,2}$ are taken in the form
\begin{equation}\label{eq:ners}
n_{e}^{(1,2)}(r)= n_{s;e}^{(1,2)}(r)+n_{ion;e}^{(1,2)}(r),
\end{equation}
where $n_{ion;e}^{(1,2)}(r)$ describes the distribution of
electrons, whose presence is caused by the electron-ion correlation,
and is obtained by the extrapolation of the second member in Eq.
(\ref{eq:22}) into the considered regions. The  member
$n_{s;e}^{(1,2)}(r)$ has to satisfy the conditions
\begin{equation}\label{eq:ners2}
\int_{0}^{l_{s;1,2}}{[n_{s;e}^{(1,2)}(r)-N_{e}\cdot (1-\alpha_{e;1,2})] \cdot 4\pi
r^{2}}dr = \alpha_{e;1,2}\cdot Z_{1,2},
\end{equation}
\begin{equation}\label{eq:bc}
n_{s;e}^{(1,2)}(r=l_{s;1,2}) = N_{e}\cdot (1 - \alpha_{e;1,2}),
\end{equation}
where Eq. (\ref{eq:ners2}) represents such a generalization of the corresponding condition
from Part 2, which is adequate in both $Z_{1}\ne Z_{2}$ and $Z_{1}=Z_{2}$
cases (see Section~\ref{sec:par}).

The necessary expressions for $n_{s;e}^{(1,2)}(r)$ in the regions $0<r<l_{s;1,2}$
are obtained by means of the data from Part 2, by appropriate replacement
of the origin designations. They are given below by Eqs. (\ref{eq:45}) and
(\ref{eq:46}). As one can see, these expressions provide automatical satisfaction of the
boundary conditions (\ref{eq:bc}). This is important since in this case Eqs.
(\ref{eq:ners2}) and (\ref{eq:ners2}) are enough for determination of the
parameters $\alpha_{e;1,2}$ and  $l_{s;1,2}$.

\subsection{The ion densities}
{\bf  The region of large $r$.} From Eqs. (\ref{eq:1}),
(\ref{eq:2}) and (\ref{eq:22}) follows the general relation for the charge
densities in the region of middle and large $r$, namely
\begin{equation}\label{eq:rho}
\rho^{(1,2)}(r) = e\cdot [Z_{1}\cdot(n_{1}^{(1,2)}(r) - N_{1}) +
Z_{2}\cdot(n_{2}^{(1,2)}(r) - N_{2})] \cdot (1 - \alpha_{e;1,2}).
\end{equation}
One can also see that this relation is insufficient for further investigation.
Namely, it is not possible to satisfy the conditions (\ref{eq:three}) if
$[n_{2,1}^{(1,2)}(r)- N_{2,1}]$ in the whole regions $l_{s;1,2} < r < \infty$ are
described by means the equations similar to Eq. (\ref{eq:121}). Therefore a separate
procedure for determination of $n_{2,1}^{(1,2)}(r)$ had to be developed here.

A direction of that development was chosen which is based
on a physically justified assumption. Namely, we take into account that
the rules found in the case of two kinds of positively charged
ions should not change drastically if one kind of ions is replaced by a different one.
In accordance with this, the next step should be choosing a system with
two kinds of positively charged particles (which could be easily described
at least qualitatively) and finding the expression for the corresponding
charge densities for the region of large and middle $r$. After that it only remains for us
to switch to the considered system, taking into account that in such
procedures some parameters should appear which could be determined based on the
corresponding physical conditions.  This fact will lower the possible error
in determination of the required charge density. Besides, we keep in mind
that, similarly to the above case of the electron density,
the changes of the ion densities in Eq. (\ref{eq:rho}) are also
very small in the region of large $r$ and could be well described in linear approximation.

As for the mentioned system which could be easily described, one can see that
an obvious choice can be an imaginary system where the positively charged ions of one kind are replaced with positrons,
which are treated here as a probe system.
Certainly, instead of the electron component in that system the non-structured
negatively charged background has to be taken, in order to eliminate the electron-positron interaction from the consideration.

Because of the great amount of the necessary material, the chosen probe system is described in details
in Supplementary material A. The result of the procedure developed in this Supplementary material is the exceptionally significant
relation, given by Eq. (\ref{eq:dif}), where the parameter $\alpha_{i}$ exists
which has the meaning of the ion-ion correlation coefficient. Namely, from this
relation and Eq. (\ref{eq:rho}) the expression
\begin{equation}
\label{eq:22a} \rho^{(1,2)}(r) =Z_{1,2}e\cdot(1-\alpha_{e;1,2})\cdot(1
-\alpha_{i}) \cdot [n_{1,2}^{(1,2)}(r) - N_{1,2}],
\end{equation}
is obtained,
which provides the applicability of the procedure for determination of the charge
and ion densities developed in Part 1 and 2. In accordance with that procedure this
expression, by means Eq. (\ref{eq:121}) for $n_{1,2}^{(1,2)}(r)$ with $r_{st} =
\infty$, and Eq. (\ref{eq:6}) for $\Phi^{(1,2)}(r)$, is transformed to an
integral equation of Volterra's type:
\begin{equation}
\rho^{(1,2)}(r) =
\kappa_{as;1,2}^{2}\int\limits_{r}^{\infty} {\rho^{(1,2)}
(r')\left({\frac{1}{r} - \frac{1}{r'}} \right) r'^{2}}dr',
\end{equation}
\begin{equation}\label{eq:kappa12as}
\kappa_{as;1,2} =\kappa_{0;1,2}\cdot[(1 -
\alpha_{e;1,2})\cdot(1
-\alpha_{i})]^{1/2}, \quad
\kappa _{0;1,2} =\left[\frac{4\pi(Z_{1,2}e)^2}{\partial \mu_{1,2}/\partial
N_{1,2}}\right]^{1/2},
\end{equation}
where the lower index $"as"$ points to the fact that it is just the ion screening constant
$\kappa_{as;1,2}$ that determines the behaviour of $\rho^{(1,2)}(r)$ in the asymptotic
region of $r$. According to Part 1 and 2, the solution of this equation  can be
taken in the form
\begin{equation}
\label{eq:rhoas} \rho^{(1,2)}(r) = \rho_{1,2}\cdot
r_{b;1,2}\exp(\kappa_{as;1,2} r_{b;1,2})\cdot
\frac{e^{-\kappa_{as;1,2}r}}{r},
\end{equation}
where $\rho_{1,2}\equiv \rho^{(1,2)}(r_{b;1,2})$ is determined from
the condition of continuity of $\rho^{(1,2)}(r)$ in the points
$r = r_{b;1,2}$. Finally, by means Eqs. (\ref{eq:22a}),
(\ref{eq:dif}) and (\ref{eq:rhoas}) we obtain the expressions for
the ion densities
\begin{equation}
n_{1,2}^{(1,2)}(r) = N_{1,2} + \frac{\rho_{1,2}\cdot
r_{b;1,2}\exp(\kappa_{as;1,2}r_{b;1,2})}{Z_{1,2}e
\cdot(1-\alpha_{e;1,2})(1-\alpha_{i})}\cdot
\frac{\exp(-\kappa_{as;1,2}r)}{r},
\end{equation}
\begin{equation}\label{eq:342}
n_{2,1}^{(1,2)}(r)=N_{2,1}-\frac{\rho_{1,2}\cdot r_{b;1,2}
\exp(\kappa_{as;1,2}r_{b;1,2})\cdot\alpha_{i}} {Z_{2,1}e\cdot(1 -
\alpha_{e;1,2})(1 - \alpha_{i})}\cdot
\frac{\exp(-\kappa_{as;1,2}r)}{r},
\end{equation}
which are valid in the regions $r_{b;1,2}<r<\infty$.\\

\noindent {\bf The regions of middle and small $r$.} In accordance with
\ref{sec:mdmi} the charge densities $\rho^{(1,2)}(r)$ in the regions
$l_{s;1,2}<r<r_{b;1,2}$ can be determined directly from Eq. (\ref{eq:rho}).
Namely, by means Eq. (\ref{eq:121}) for $n_{1,2}^{(1,2)}(r)$ with
$r_{st}=r_{b;1,2}$, Eq. (\ref{eq:121ion}) for $n_{2,1}^{(1,2)}(r)$ and Eq.
(\ref{eq:6}) for $\Phi^{(1,2)}(r)$, Eq. (\ref{eq:rho}) is transformed to the
integral equation
\begin{equation}
\rho^{(1,2)}(r) = \frac{\kappa_{i}^{2}}{r}\cdot
\left[\int\limits_{r}^{r_{b;1,2}}{\rho^{(1,2)}(r')(r' - r)r'}dr' +
\frac{\rho_{1,2}\cdot(r_{b;1,2} - r)(1 +\kappa_{as;1,2}r_{b;1,2})}
{\kappa_{as;1,2}^{2}}\right] +\rho_{1,2},
\end{equation}
\begin{equation}\label{eq:kappai1}
\kappa_{i}=[(\kappa_{0;1}^2 + \kappa_{0;2}^2)\cdot
(1-\alpha_{e;1,2})]^{1/2},
\end{equation}
where $\rho_{1,2}= \lim\limits_{r \to r_{b;1,2}}\rho^{(1,2)}(r)\equiv
\rho^{(1,2)}(r_{b;1,2})$. In accordance with Eq.
(\ref{eq:relAB}) from \ref{sec:mdmS} the solutions of this equation are
given by
\begin{equation}\label{eq:rhorho}
\rho^{(1,2)}(r) = - c_{1,2}\cdot e N_{1,2}r_{b;1,2}\cdot \frac{F_{1,2}(r)}{r},
\end{equation}
where $F_{1,2}(r)$ is given below by Eq.(\ref{eq:ABCD}) with complete
expressions for the ion densities, and $c_{1,2}>0$ is the unknown parameter. From
here and Eqs. (\ref{eq:121}), (\ref{eq:rho}) and (\ref{eq:kappai1}) it follows
that the ion densities in the regions $l_{s;1,2}<r<r_{b;1,2}$ are given by
\begin{equation}
n_{1,2}^{(1,2)}(r)= N_{1,2} - \frac{c_{1,2}N_{1,2}}{Z_{1,2}(1-\alpha_{e;1,2})}
\cdot \left[(1-f_{1,2})\frac{1-d_{1,2}(1-\alpha_{i})}{1-\alpha_{i}} +
d_{1,2}r_{b;1,2}\cdot\frac{F_{1,2}(r)}{r}\right],
\end{equation}
\begin{equation}\label{eq:342a}
n_{2,1}^{(1,2)}(r) = N_{2,1} +
\frac{c_{1,2}N_{1,2}}{Z_{1,2}(1-\alpha_{e;1,2})} \cdot
\left[(1-f_{1,2})\frac{1-d_{1,2}(1-\alpha_{i})}{(1-\alpha_{i})} -
d_{2,1}r_{b;1,2}\cdot\frac{F_{1,2}(r)}{r}\right],
\end{equation}
where $f_{1,2}$ and $d_{1,2}$ are defined below by Eq. (\ref{eq:ABCD} together
with $F_{1,2}(r)$. Then, we will introduce other unknown parameters
$r_{0;1,2}^{(1,2)}$, taking $c_{1,2}$ in the form
\begin{equation}\label{eq:c12}
c_{1,2} =\frac{Z_{1,2}(1-\alpha_{e;1,2})(1-\alpha_{i})}{(1-f_{1,2})[1-d_{1,2}
(1-\alpha_{i})]+ (r_{b;1,2}/r_{0;1,2}^{(1,2)})d_{1,2}(1-\alpha_{i})\cdot
F_{1,2}(r_{0;1,2}^{(1,2)})},
\end{equation}
which provides that $n_{1,2}^{(1,2)}(r = r_{0;1,2}^{(1,2)})\equiv 0$.

Here we will take into account the fact that the behavior of the ions of
both considered kinds in the regions of small $r$ has to be similar
to that of positively charged ions in the vicinity of the
positive charged probe particle in the single- and two-component
systems. Therefore we will determine the ion densities in the
regions $0 < r< l_{s;1,2}$ using the procedure of extrapolation,
which is described and discussed in detail in Part 1 and 2. This
means the following: extrapolation of Eqs. (\ref{eq:342a}) and
(\ref{eq:c12}) for $n_{1,2}^{(1,2)}(r)$ to the points
$r_{0;1,2}^{(1,2)}$ and for $n_{2,1}^{(1,2)}(r)$ to the points
$r_{0;2,1}^{(1,2)}$ such that $n_{2,1}^{(1,2)}(r_{0;2,1}^{(1,2)}) =
0$, and after that - extension to the point $r=0$ by zero.

\section{(i1) and (i2) cases:  complete expressions}
\label{sec:i12}

\subsection{ The ion densities}
\label{sec:i12a}

The complete expressions for the ion densities follow just
from the above mentioned and Eqs. (\ref{eq:342}), (\ref{eq:rhorho}),
(\ref{eq:342a}), (\ref{eq:rhorho}) and (\ref{eq:c12}). Due to their importance
they are presented separately for {\bf the case $(i1)$} in the form
\begin{equation}\label{eq:n1infty1}
n_{1}^{(1)} (r) = N_{1}\cdot\left\{ {\begin{array}{*{30}l} 0, & 0 <
r \le r_{0;1}^{(1)} \\
\displaystyle{ 1 - A_{1} - B_{1}d_{1}r_{b;1}\cdot \frac {F_{1}(r)}{r}, }
& r_{0;1}^{(1)} < r \le r_{b;1}, \\
\displaystyle{1 -
C_{1}r_{b;1}\cdot \frac{e^{-\kappa_{as;1}(r-r_{b;1})}}{r},}
\hfill & {r_{b;1} < r < \infty ,} \hfill \\
\end{array} } \right.
\end{equation}
\begin{equation}\label{eq:n2infty1}
n_{2}^{(1)}(r) = N_{2}\cdot\left\{ \begin{array}{*{30}l} 0, & 0 < r
\le r_{0;2}^{(1)} \\
\displaystyle{1 + \frac{N_{1}Z_{1}}{N_{2}Z_{2}}\cdot A_{1} -
\frac{N_{1}Z_{1}}{N_{2}Z_{2}}\cdot B_{1}d_{2}}r_{b;1}\cdot \frac {F_{1}(r)}{r},
\hfill & {r_{0;2}^{(1)} < r \le r_{b;1}} , \hfill \\
\displaystyle{1+\frac{N_{1}Z_{1}}{N_{2}Z_{2}}\cdot C_{1}\alpha_{i}r_{b;1}\cdot
\frac{e^{-\kappa_{as;1}(r-r_{b;1})}}{r}},
\hfill & {r_{b;1} < r < \infty ,} \hfill \\
\end{array} \right.
\end{equation}
and separately for {\bf the case $(i2)$} in the similar form
\begin{equation}\label{eq:n2infty2}
n_{2}^{(2)} (r) = N_{2}\cdot\left\{\begin{array}{*{30}l}
0, & 0 < r \le r_{0;2}^{(2)} \\
\displaystyle{ 1 - A_{2} - B_{2}d_{2}r_{b;2}\cdot \frac {F_{2}(r)}{r},}
\hfill & {r_{0;2}^{(2)} < r \le r_{b;2} ,} \hfill \\
\displaystyle{1 -
C_{2}r_{b;2}\cdot \frac{e^{-\kappa_{as;2}(r-r_{b;2})}}{r},}
\hfill & r_{b;2} < r < \infty , \hfill \\
\end{array}  \right.
\end{equation}
\begin{equation}\label{eq:n1infty2}
n_{1}^{(2)}(r) = N_{1}\cdot\left\{ \begin{array}{*{30}l}
0, & 0 < r \le r_{0;1}^{(2)} \\
\displaystyle{1 + \frac{N_{2}Z_{2}}{N_{1}Z_{1}}\cdot A_{2} -
\frac{N_{2}Z_{2}}{N_{1}Z_{1}}\cdot B_{2}d_{1}r_{b;2}\cdot \frac {F_{2}(r)}{r}},
\hfill & {r_{0;1}^{(2)} < r \le r_{b;2}} , \hfill \\
\displaystyle{1+\frac{N_{2}Z_{2}}{N_{1}Z_{1}}\cdot C_{2}\alpha_{i}r_{b;2}\cdot
\frac{e^{-\kappa_{as;2}(r-r_{b;2})}}{r}},
\hfill & {r_{b;2} < r < \infty ,} \hfill \\
\end{array} \right.
\end{equation}
where the functions $F_{1,2}(r)$ and the coefficients $f_{1,2}$, $A_{1,2}$,
$B_{1,2}$, $C_{1,2}$ and $d_{1,2}$ are given by the relations
\begin{eqnarray}
F_{1,2}(r)\equiv e^{\kappa_{i}(r_{b;1,2}-r)}- f_{1,2} \cdot
e^{-\kappa_{i}(r_{b;1,2}-r)}, \\ \nonumber
f_{1,2} = \frac{(1 +
\kappa_{as;1,2}r_{b;1,2})\cdot\kappa_{i}^2 -
(1 + \kappa_{i}r_{b;1,2})\cdot\kappa_{as;1,2}^2}{(1 +
\kappa_{as;1,2}r_{b;1,2})\cdot \kappa_{i}^2 - (1 -
\kappa_{i}r_{b;1,2})\cdot\kappa_{as;1,2}^2},
\end{eqnarray}
\begin{equation}
A_{1,2}= [1-d_{1,2}(1-\alpha_{i})]\cdot C_{1,2}, \qquad
B_{1,2}=\frac{1-\alpha_{i}}{1-f_{1,2}}\cdot C_{1,2}
\end{equation}
\begin{equation}\label{eq:ABCD}
C_{1,2}=\left[1-d_{1,2}(1-\alpha_{i})\left(1-\frac{r_{b;1,2}}{1-f_{1,2}}
\cdot\frac {F_{1,2}(r_{0;1,2}^{(1,2)})}{r_{0;1,2}^{(1,2)}}
\right)\right]^{-1},
\quad d_{1,2} =
\frac{\kappa_{0;1,2}^{2}}{\kappa_{0;1}^{2}+\kappa_{0;2}^{2}},
\end{equation}
and the screening constants $\kappa_{as;1,2}$, $\kappa_{0;1,2}$ and $\kappa_{i}$ -
by Eqs. (\ref{eq:kappa12as}) and (\ref{eq:kappai1}).

\subsection{ The electron densities} The complete expressions for the electron
densities are obtained from Eqs. (\ref{eq:22}), (\ref{eq:ners}) and
(\ref{eq:bc}), and are presented here in the form
\begin{equation}\label{eq:ne12}
n_{e}^{(1,2)} (r) = \alpha_{e;1,2}\cdot[Z_{1}\cdot n_{1}^{(1,2)}(r)+
Z_{2}\cdot n_{2}^{(1,2)}(r)] + \left\{ {{\begin{array}{*{20}c}
\displaystyle{n_{s;e}^{(1,2)}(r) , } \hfill & {0 < r \le l_{s;1,2}
,}
\hfill \\
\displaystyle{N_{e}\cdot (1 - \alpha_{e;1,2}),}
\hfill & {l_{s;1,2} < r < \infty ,} \hfill \\
\end{array} }} \right.
\end{equation}
where the ion densities $n_{1}^{(1,2)}(r)$ and $n_{2}^{(1,2)}(r)$
are given by Eqs. (\ref{eq:n1infty1}) - (\ref{eq:n2infty2}). The number
$n_{s;e}^{(1,2)}(r)$, in accordance with the above, is given by the relations
\begin{equation}\label{eq:45}
n_{s;e}^{(1,2)} (r) = N_{e}\cdot l_{s;1,2}\frac{a_{1,2}
\cdot e^{- \kappa_{0;e}r} + b_{1,2} \cdot e^{\kappa_{0;e}r}}{r},
\qquad 0 < r < l_{s;1,2},
\end{equation}
\begin{equation}
a_{1,2} = \frac{1 - \alpha_{e;1,2} - \frac{1}{3}x_{l;1,2}^{2} \cdot e^{x_{l;1,2}}}
{e^{- x_{l;1,2}}- e^{x_{l;1,2}}}, \qquad
b_{1,2} = - \frac{1 - \alpha_{e;1,2} -\frac{1}{3}x_{l;1,2}^{2}\cdot e^{-x_{l;1,2}}}
{e^{- x_{l;1,2}} - e^{x_{l;1,2}}},
\end{equation}
\begin{equation}\label{eq:46}
x_{l;1,2}=\kappa _{0;e}\cdot l_{s;1,2}, \qquad
\kappa _{0;e} = \left[\frac {4\pi e^{2}}{\partial \mu_{e}/\partial
N_{e}}\right]^{1/2}.
\end{equation}
One can see that these expressions provide automatical satisfaction of the
boundary conditions (\ref{eq:bc}). This is important since in this case Eqs.
(\ref{eq:1c}) and (\ref{eq:ners2}) are enough for determination of the
parameters $\alpha_{e;1,2}$ and  $l_{s;1,2}$. Let us note that these parameters, as
well as $r_{b;1,2}$, $r_{0;1}^{(1,2)}$, $r_{0;2}^{(1,2)}$ and $r_{0;1}^{(1,2)}$,
are determined as it is described in Section \ref{sec:par}.

\section{The presented method: the case (e)}
\label{sec:mdme}

\subsection{The partial expressions}
\label{sec:mdme1}

{\bf The ion and electron densities: the region of large $r$.}
On the basis of discussion from Part 2 we will take into account that in the auxiliary
system $S_{a}^{(e)}$ only the electron component can be considered
an ideal gas, while the ion components can be treated as parts of some
stationary, but non-homogenous positively charged background. This is
a consequence of the fact that a fixed negatively charged probe
particle, which certainly makes physical sense in the case of some
electrolyte, cannot be the representative of a free electron (light
particle), surrounded by positively charged ions (heavy
particles), particulary for the fact that ions were already treated as
immobile with respect to the electrons. In the previous Section it
was mentioned that similar approximations have very often been used
in the physics of plasma. As it is well known, the reason for this is
very short time for which a free electron stays in the vicinity of any ion, which
makes it possible to neglect the influence of the electron-ion
interaction on the ion positions. Therefore we must not
describe the ion components in the system $S_{a}^{(e)}$ in a
way similar to that in the systems $S_{a}^{(1,2)}$. Namely, according to Part
2, the ion distributions in the system $S_{a}^{(e)}$ as a matter of fact have
to reflect the properties of the electron distributions with respect to
the positively charged ions in plasma, i.e. in the considered case -
the properties of the already described electron distributions in
the systems $S_{a}^{(1,2)}$, including the probe-particle
self-spheres. This means that in the system $S_{a}^{(e)}$
the positive charge density has to be determined
$\rho_{i}^{(e)}(r)=Z_{1}e\cdot n_{1}^{(e)}(r)+Z_{2}e\cdot
n_{2}^{(e)}(r)$, where the functions $n_{1}^{(e)}(r)$ and
$n_{2}^{(e)}(r)$ can be treated as ion densities, although they
have only one task: to make possible treating the probe particle potential
energy in the system $S_{a}^{(e)}$ as a good
approximation of the mean electron potential energy in the system
$S_{in}$.

Since the adequate procedure of determination of $\rho_{i}$ in
the system corresponding to $S_{a}^{(e)}$, which was developed in
Part 2, allows generalization to the systems with more positively
charged ion components, it is just that procedure that is used here,
certainly with the necessary modifications. So, the probabilities of
changes of the partial charge densities $Z_{1,2}e \cdot
n_{1,2}^{(e)}(r)$ with the corresponding changes of $n_{e}^{(e)}(r)$
are not equal to $\alpha_{e;1,2}$, but to $\alpha_{e;1,2}\cdot
p_{1,2}$. Consequently, Eqs. (78) and (35) from Part 2 are
transformed to the form
\begin{equation}\label{eq:72a}
\alpha_{e;1,2} \cdot p_{1,2} = Z_{1,2} \cdot \frac{N_{1,2} - n_{1,2}^{(e)}(r)}
{N_{e}-n_{e}^{(e)}(r)}, \quad n_{1,2}^{(e)}(r) = N_{1,2}\cdot(1-\alpha_{e;1,2})+
\frac{\alpha_{e;1,2}}{Z_{1,2}} \cdot p_{1,2} \cdot n_{e}^{(e)}(r).
\end{equation}
From here and Eq. (\ref{eq:2}) the relation for
$\rho^{(e)}(r)$ follows, namely
\begin{equation}\label{eq:relrho}
\rho^{(e)}(r)=-e\cdot [n_{e}^{(e)}(r)-N_{e}]\cdot(1-\alpha_{e;1,2}),
\quad r >l_{e;max}\equiv max (l_{e;1},l_{e;2}).
\end{equation}
Then, from this relation and Eq. (\ref{eq:121}) for $n_{e}^{(e)}(r)$, with
$r_{st} =\infty$ and $\Phi^{(e)}(r)$ given by Eq. (51) from Part 1,
the necessary integral equation is obtained:
\begin{equation}\label{eq:rhoe}
\rho^{(e)}(r)= \kappa_{e}^{2}\int\limits_{r}^{\infty}
{\rho^{(e)}(r')\left({\frac{1}{r} - \frac{1}{r'}} \right) r'^{2}}dr',
\qquad \kappa_{e} = \kappa_{0;e}\cdot(1 -
\alpha_{e;1,2})^{1/2},
\end{equation}
where $\kappa_{e}$ is the electron screening constant. Since it is similar to Eq.
(\ref{eq:kappa12as}), the charge and electron densities in the region $l_{e;max}
< r <\infty$ can be taken in the known form
\begin{equation}\label{eq:nease}
\rho^{(e)}(r)=-eN_{e}(1-\alpha_{e;1,2})r_{0;e}\cdot
\frac{e^{-\kappa_{e}(r_{0;e}-r)}}{r}, \quad
n_{e}^{(e)}(r) = N_{e} - N_{e}r_{0;e}(1 - \alpha)\cdot
\frac{e^{-\kappa_{e}(r_{0;e}-r)}}{r},
\end{equation}
which provides that $n_{e}^{(e)}(r=r_{0;e})\equiv 0$, where
$r_{0;e}$ is the unknown parameter. The electron density in the
region $0<r<l_{e;max}$ will be determined by means of the procedure
similar to the one which is used in the previous Section for the ion
components in the region of small $r$.\\

\noindent {\bf The ion densities: the region of small $r$.} In accordance with Part 2 the
ion densities $n_{1,2}^{(e)}$ inside the probe particle self-sphere are taken
here in the form
\begin{equation}\label{eq:72b}
n_{1,2}^{(e)}(r) = n_{s;1,2}^{(e)}(r) + n_{el;1,2}^{(e)}(r), \qquad
0 < r < r_{e},
\end{equation}
where $n_{el;1,2}^{(e)}(r)$ describes the distribution of the ions,
whose presence is caused by the electron-ion correlation, and is
obtained by extrapolation of the second member in Eq.
(\ref{eq:72a}) into the regions $0< r < l_{e;1,2}$. The member
$n_{s;1,2}^{(e)}(r)$ has to satisfy the conditions
\begin{equation}\label{eq:bci}
n_{s;1,2}^{(e)}(r=l_{e;1,2}) = N_{1,2}\cdot (1 - \alpha_{e;1,2}),
\end{equation}
\begin{equation}\label{eq:bciint}
\int_{0}^{l_{e;1,2}}{[n_{s;1,2}^{(e)}(r)-N_{1,2}\cdot (1-\alpha_{e;1,2})]\cdot 4\pi
r^{2}}dr= \frac{\alpha_{e;1,2} p_{1,2}}{Z_{1,2}},
\end{equation}
where the second one represents a generalization of the
condition (41) from Part 2 which is applicable in both $Z_{1}\ne
Z_{2}$ and $Z_{1}=Z_{2}$ cases (see Section~\ref{sec:par}). The
introduction of two characteristic lengths $l_{e;1,2}$ instead of one
$r_{e}$ is caused by the presence of two ion components. The
expressions for $n_{s;1,2}^{(e)}(r)$ are obtained here by
corresponding replacement of the origin designations of Eq. (43) from
Part 2, and are given below by Eq. (\ref{eq:45e}).

\subsection{The electron and ion densities: complete expressions}
\label{sec:mdme1}

Here we can repeat the procedures from Part 2 word-for-word. So, starting
from Eq. (\ref{eq:rhoe}), we obtain the expression for the electron density
\begin{equation}\label{eq:68}
n_{e}^{(e)} (r) = \left\{ {{\begin{array}{*{20}c} \displaystyle{
{N_{e} - N_{e} r_{0;e} \cdot \exp (\kappa_{e} r_{0;e}) \cdot
\frac{\exp ( - \kappa_{e}r)}{r},}} \hfill & {r_{0;e} < r < \infty ,}
\hfill \\ {0,} \hfill & {0 < r \le r_{0;e} ,} \hfill \\
\end{array} }} \right.
\end{equation}
which determines $n_{e}^{(e)}(r)$ in the whole region $0<r<\infty$. Then, in
accordance with the previous subsection, we obtain the expressions
for the ion densities in the same regions by means of Eqs. (\ref{eq:72a}) and
(\ref{eq:72b}), the conditions (\ref{eq:bci}) and (\ref{eq:bciint}), and Eq. (43)
from Part 2. The obtained expressions are given here by the relations
\begin{equation}\label{eq:76}
n_{1,2}^{(e)}(r)=\frac{\alpha_{e;1,2}p_{1,2}}{Z_{1,2}}\cdot
n_{e}^{(e)}(r)+ \left\{ {{\begin{array}{*{20}c} \displaystyle{
{N_{1,2}(1-\alpha_{e;1,2}),}}
\hfill & {l_{e;1,2} < r < \infty ,} \hfill \\
\displaystyle{{n_{s;1,2}^{(e)}(r),}}
\hfill & {0 < r \le l_{e;1,2} ,} \hfill \\
\end{array} }} \right.
\end{equation}
\begin{equation}\label{eq:55}
n_{s;1,2}^{(e)}(r) =  N_{1,2} r_{e} \frac
{a_{1,2}\cdot e^{-\frac{x_{l;1,2}\cdot r}{l_{e;1,2}}}+
b_{1,2}\cdot e^{\frac{x_{l;1,2}\cdot r}{l_{e;1,2}}}}{r}, \qquad 0 <r
\le l_{e;1,2},
\end{equation}
\begin{equation}\label{eq:45e}
l_{e;1,2} = r_{e} \cdot \frac {l_{s;1,2}}{r_{s;1,2}}
\end{equation}
where $a_{1,2}$ and $b_{1,2}$ are given by Eq. (\ref{eq:46}). One
can see that in such a form $n_{s;1,2}^{(e)}(r)$ automatically
satisfy the boundary condition (\ref{eq:bci}), as well as the
condition (\ref{eq:bciint}). Let us note that the the ratios
$l_{s;1,2}/r_{s;1,2}$ and the characteristic length $r_{0;e}$ are
determined as it is described in the next Section.

\section{Determination of the parameters}
\label{sec:par}

The parameters $\alpha_{e;1,2}$ and $l_{s;1,2}$ are determined
differently in the cases $Z_{1}=Z_{2}= Z_{i}$ and $Z_{1}\ne
Z_{2}$. Namely, within the used procedure the first case, where
$r_{s;1} = r_{s;2}\equiv r_{s;i}$, is equivalent (from the point of view of
determination of $\alpha_{e;1,2}$) to the case of the two-component
plasma with the same $Z_{i}$, $N_{e}$ and $T_{e}$. Consequently, in
the case $Z_{1,2}= Z_{i}$ the following relations
\begin{equation}\label{eq:alp1}
l_{s;1,2}=r_{s;i}, \qquad \alpha_{e;1,2}=\alpha(x_{s;i}), \quad
x_{s;i}\equiv \kappa_{0;e}\cdot r_{s;i},
\end{equation}
are valid, where in accordance with Part 2, $\alpha(x)$ is defined by
\begin{equation}
\label{eq:alphax}
\alpha(x)=1 - \frac{\frac{2}{3} x^{3}}{(1+x)\cdot e^{-x} - (1-x)\cdot e^{x}}.
\end{equation}
However, in other case we have it that $r_{s;1}\ne r_{s;2}$ and,
consequently, $l_{s;1}\ne l_{s;2}$. Therefore, in the case
$Z_{1} \ne Z_{2}$ the parameters $\alpha_{e;1,2}$ and $l_{s;1,2}$ have
to be determined together from Eqs. (\ref{eq:ners2}) and
(\ref{eq:bc}). For that purpose these equations have to be
transformed first, by means Eqs. (\ref{eq:45}) and (\ref{eq:46}),
to the form
\begin{equation}\label{eq:alpha}
\frac{l_{s;1,2}^{3}[(1+x_{l;1,2})e^{-x_{l;1,2}}-
(1-x_{l;1,2})e^{x_{l;1,2}}-\frac{2}{3}x_{l;1,2}^{3}]}
{l_{s;1,2}^{3}[(1+x_{l;1,2})e^{-x_{l;1,2}}-(1-x_{l;1,2})e^{x_{l;1,2}}]-
\frac{2}{3}x_{l;1,2}^{3}\frac{e^{-x_{l;1,2}}-e^{x_{l;1,2}}}{2x_{l;1,2}}
(r_{s;1,2}^{3} -l_{s;1,2}^{3})} = \alpha_{e;1,2},
\end{equation}
\begin{equation}\label{eq:ls}
p_{1}\cdot (l_{s;1}/r_{s;1})^{3} + p_{2}\cdot
(l_{s;2}/r_{s;2})^{3} = 1,
\end{equation}
where $x_{l;1,2}$ is given by Eq. (\ref{eq:46}). Then, the variables
$x_{l;1,2}$ in Eq. (\ref{eq:alpha}) have to be replaced with
$x_{s;1,2}\cdot l(s;1,2)/ r_{s;1,2}$, where $x_{s;1,2}= k_{0;e}\cdot
r_{s;1,2}$. After that the ratios $l_{s:1,2}/r_{s;1,2}$ can be used
as variables in both equations. Obtaining their solution
is aided by the relations
\begin{equation}\label{eq:alp12}
\alpha_{e;1,2} \cong \alpha(x_{s}=x_{s;1})\cdot p_{1} +
\alpha(x_{s}=x_{s;2})\cdot p_{2}, \quad |l_{s:1,2}/r_{s;1,2} - 1| <<1,
\end{equation}
which are established by direct calculations. Also, it is useful
to keep in mind the fact that the left side of Eq. (\ref{eq:alpha})
approaches to $\alpha(x_{s;i})$ if $l(s;1,2)$ approaches to
$r_{s;i}$, and that it is always $l_{s;1} > r_{s;1}$ and $l_{s;2}<
r_{s;2}$ if $Z_{1}<Z_{2}$.

It is important that the electron-ion correlation coefficient
$\alpha_{e;1,2}$ and the characteristic lengths $l_{s;1,2}$ are
determined, as it is described below, independently of all other
parameters. Namely, that is why the existing conditions are
sufficient for determination of the characteristic lengths
$r_{0;1,2,e}^{(1,2,e)}$, $r_{0;2,1}^{(1,2)}$ and $r_{b;1,2}$, and
the ion-ion correlation coefficient $\alpha_{i}$.

So, $r_{0;1,2}^{(1,2)}$ are determined as functions of
$\alpha_{i}$ and $r_{b;1,2}$ from Eq. (\ref{eq:13}) for
$n_{1,2}^{(1,2)}$, and $r_{0;2,1}^{(1,2)}$ - as the roots of the
equations $n_{2,1}^{(1,2)}(r)= 0$ which can be find in the interval $[0,r_{b;1,2}]$.
The very important parameters $r_{b;1,2}$, i.e. the distances from the point $O$ at which
the manner of describing the ion densities changes, are determined
from Eq. (\ref{eq:three}) for $n_{2,1}^{(1,2)}$ through a procedure where it is taken that
\begin{equation} \label{eq:rb12}
r_{b;1,2} = r_{s;1,2} \cdot (1+\eta_{1,2}), \quad 0 < \eta_{1,2} \le \eta_{max;1,2}, \quad \eta_{max;1,2} \gg 1
\end{equation}
where $\eta_{1,2}$ are new parameters which are used in the calculations
in such a way that they vary with the small steps $\Delta \eta_{1,2}=1/K_{1,2}$
where $K_{1,2}\gg 1$.
As the results of this procedure we get the values of the parameters $r_{b;1,2}$,
i.e. the main one of the considered characteristic lengths,
which corresponds to the current value of the ion-ion correlation coefficient $\alpha_{i}$. The final value
of this coefficient itself is determined through a procedure which implies scanning
$\alpha_{i}$ with a very small step in the interval from 0 to 1,
and examining at each step whether the equation
\begin{equation}\label{eq:addc}
r_{0;1}^{(2)} - r_{0;2}^{(1)} = 0,
\end{equation}
is satisfied,
which provides the physical meaning of the obtained solutions. The whole procedure ends when
the equation is satisfied.

Then, $r_{0;e}$ is determined from Eq. (\ref{eq:13}) for $n_{e}^{(e)}(r)$, as in
Part 1 and 2. Therefore it can be presented in two equivalent forms, namely
\begin{equation} \label{eq:70}
r_{0;e} = \frac{(1+x^{3})^{\frac{1}{3}}-1}{x}\cdot r_{e}\equiv
\gamma_{s;e}(x) \cdot r_{e},
\qquad r_{0;e}=[(1+x^{3})^{\frac{1}{3}}-1]\cdot r_{\kappa;e}\equiv
\gamma_{\kappa;e}(x)\cdot r_{\kappa;e},
\end{equation}
where $x=\kappa_{e}r_{e}$, $r_{\kappa;e}\equiv 1/\kappa_{e}$, and
the coefficients $\gamma_{s;e}(x)$ and $\gamma_{\kappa;e}(x)$ are
connected with the electron non-ideality parameters
$\Gamma_{e}=e^{2}/(kT_{e}r_{e})$ and $\gamma_{e} =
e^{2}/(kT_{e}r_{\kappa;e})$ as it is described in Part 3.

Finally, let us note that the partial electron and ion densities
$n_{s;e}^{(1,2)}$ and $n_{s;1,2}^{(e)}$, due to the structure of
the equation for coefficients $a_{1,2}$ and $b_{1,2}$, \ref{eq:46}, can
be determined by Eqs. (\ref{eq:45}), (\ref{eq:46}) and
(\ref{eq:45e}) in both $Z_{1,2} = Z_{i}$ and $Z_{1}\ne Z_{2}$ cases.
Therefore it is necessary to take
the corresponding values of $\alpha_{e;1,2}$ and $l_{s;1,2}$ only in these expressions, e.g.
$l_{s;1,2}=r_{s;i}$ if $Z_{1,2}=Z_{i}$.

The behaviour of the characteristic length $l_{s;1,2}$ and electron-ion correlation coefficients
$\alpha_{e,1,2}$ is illustrated by Tabs. 1 and 2 in the online material, while
the behaviour of the parameters $r_{b;1,2}$ and ion-ion correlation
coefficients $\alpha_{i}$ is shown there in Tabs. 3 and 4.

These tables cover the regions of $N_e$ from $10^{16}$ cm$^{-3}$ to $10^{19}$ cm$^{-3}$
for $T=3\cdot 10^{4}$ K.
They show that the values of all parameters are within the expected boundaries.
So from these tables one can see that mostly $l_{s;1,2}\approx r_{s;1,2}$,
electron-ion correlation coefficient $l_{e;1,2}\approx 1$, $r_{b;1,2}\sim r_{s;1,2}$
and ion-ion correlation coefficient $\alpha_{i}\approx 1$; a significant increase
in the correlation coefficients values $\alpha_{e,1,2}$ and $\alpha_{i}$
is registered in the region of extremely high electron density ($N_e \approx 10^{19}$ cm$^{-3}$).
These tables cover the region of electron densities $ 10^{16}$ cm$^{-3}$ $\le N_e \le 10^{20}$ cm$^{-3}$
for $T=3\cdot 10^{4}$ K.

Remark: at this point all the calculations are performed for the case  $T_{i}=T_{e}=T$.

\section{Results and discussions}
\label{sec:pso}

\subsection{The properties of the obtained solutions}
\label{sec:pso1}

One of the most important results of this research is the performed classification
of the considered systems from the point of view of application of the above
described additional conditions:
\begin{itemize}
  \item[--] the systems of the "closed" type, which are investigated here, and which are described by means of the mentioned additional conditions;
  \item[--] the systems of the "open" type, where those conditions are fully neglected.
\end{itemize}
It is just for the systems of the "closed" type that the method of describing the inner electrostatic screening is developed in this work.
As a continuation of our previous research (Part 1 and Part 2), as the systems of the "closed" type in this work
fully ionized electron ion plasmas are chosen with the positive ion charges of two different kinds. Such choice if especially important since
increasing the number of ion components further would not cause appearance of any new phenomena.
The obtained results are discussed in this section, including the results of comparison with the systems of the "open" type,

\begin{figure}[h!]
\begin{center}\includegraphics[width=0.8\columnwidth,
height=0.6\columnwidth]{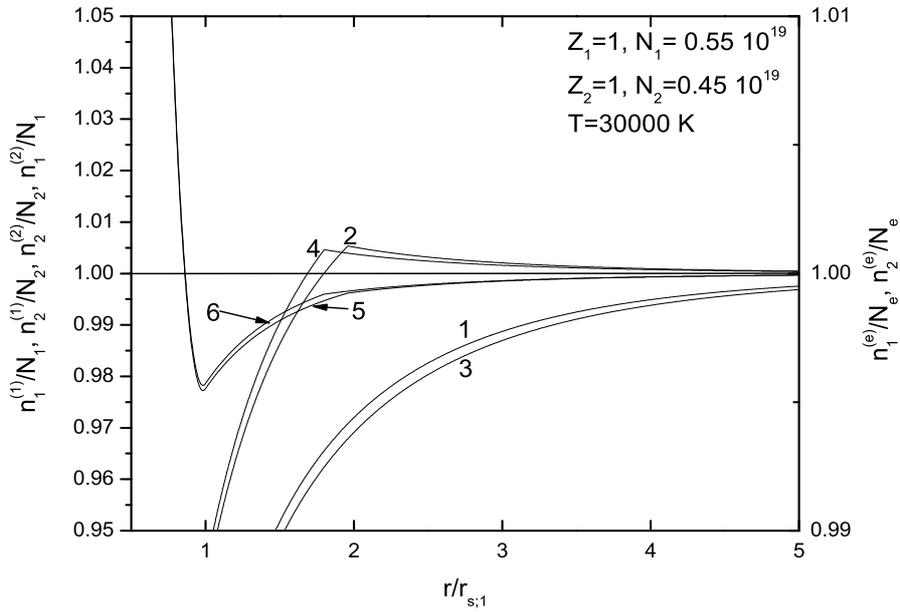}
\caption{The behavior of
reduced densities $n_{1}^{(1)} (r) / N_{1}$ (in figure curve marked with 1), $n_{2}^{(1)} (r) / N_{2}$ (curve marked with 2),
$n_{2}^{(2)} (r) / N_{2}$ (curve marked with 3) and $n_{1}^{(2)} (r) / N_{1}$  (curve marked with 4) in the case $Z_{1}=1$, $Z_{2}=1$ and $T_{i}=T_{e}=T$,
where $T=30000K$.}
\label{fig:htbp11}
\end{center}
\end{figure}
\begin{figure}[h!]
\begin{center}\includegraphics[width=0.8\columnwidth,
height=0.6\columnwidth]{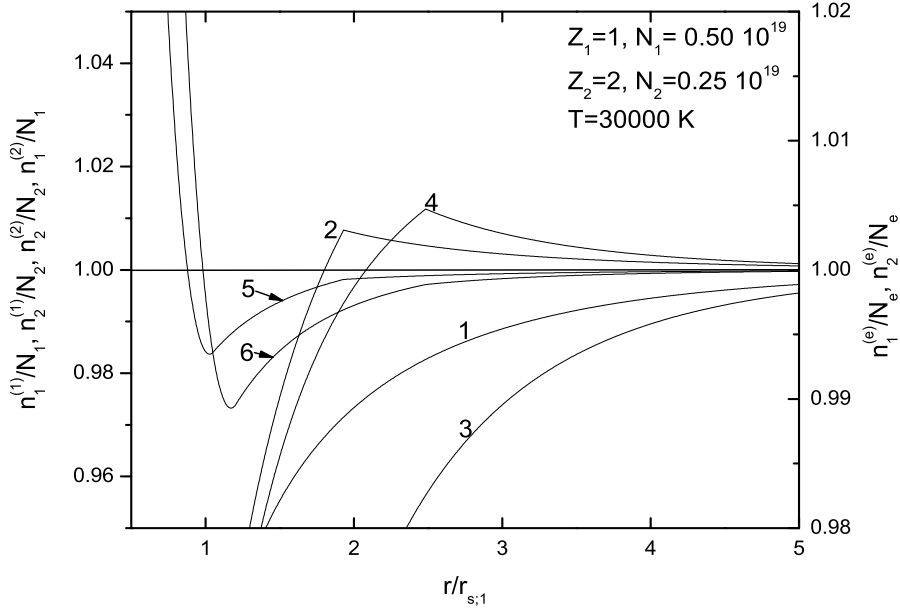}
\caption{The behavior of
reduced densities $n_{1}^{(1)} (r) / N_{1}$ (in figure curve marked with 1), $n_{2}^{(1)} (r) / N_{2}$ (curve marked with 2),
$n_{2}^{(2)} (r) / N_{2}$ (curve marked with 3) and $n_{1}^{(2)} (r) / N_{1}$  (curve marked with 4) in the case $Z_{1}=1$, $Z_{2}=2$ and $T_{i}=T_{e}=T$,
where $T=30000K$.}
\label{fig:htbp113}
\end{center}
\end{figure}
One can see that the procedures of obtaining the Eqs. (\ref{eq:n1infty1})
- (\ref{eq:46}) and Eqs. (\ref{eq:68}) - (\ref{eq:45e}), as well as the values
of the existing parameters, provide that these expressions: are self-consistent;
satisfy all the conditions from Section \ref{sec:tas}, including Eq.
(\ref{eq:three}) and Eq. (\ref{eq:addc}); can be applied not only to the
classical, but also to the quantum-mechanical systems (see Part 2.), including here
the plasmas of higher non-ideality.
Since the presented expressions do not contain the particle masses they can be used also for
describing some other systems (the corresponding electrolytes and dusty
plasmas). The behaviour of the ion and electron densities is illustrated by Fig's.
\ref{fig:htbp11} and \ref{fig:htbp113} on the examples of the cases $(i1)$ and $(i2)$
for $Z_{2}\ne Z_{1}$ and $Z_{2}=Z_{1}$ respectively.

Since Eq. (\ref{eq:45}) and Eq.(\ref{eq:55}) show that the solutions
$n_{e}^{(1,2)}(r)$ and $n_{i;1,2}^{(i)}(r)$
are singular in the point $r = 0$, it is useful to note that the
existence of singularities in model solutions is fully acceptable,
if it does not have other non-physical consequences. Such solutions are
well known in physics: it is enough to mention, for example,
Thomas-Fermi's models of electron shells of heavy atoms
(\cite{tho26,fer28}; see also \cite{gom50}), which are used in plasma
research until now (see e.g. \cite{men02}).

One more comment is due with respect to the figures \ref{fig:htbp11} and \ref{fig:htbp113}.
Namely they show that the ion densities $n_{1,2}^{(2,1)}(r)$ in the points $r_{b;1,2}$ do not keep
the smutting.  It makes it actually impossible that within
any similar ion-ion correlation model there be any other equations which would be better then Eq. (\ref{eq:121ion}).

All potential applications of the obtained solutions are connected with the
charge density $\rho^{(1,2,e)}(r)$, which in general is given by Eq. (\ref{eq:2})
and described in detail in Supplementary material C. Namely, if $\rho^{(1,2,e)}(r)$ is known
then by means of Eq. (\ref{eq:6}) the electrostatic potential
$\Phi^{(1,2,e)}(r)$ can be determined, and by means of Eq. (\ref{eq:5}) - the potential $\varphi^{(1,2,e)}$ at the point $O$.
Finally, by means of the potential $\varphi^{(1,2,e)}$  the probe particle
potential energies given by Eq.(\ref{eq8}) are determined.

Except for the potential $\Phi^{(1,2,e)}(r)$ and $\varphi^{(1,2,e)}$
the systems $S_{a}^{(1,2,e)}$ are certainly characterized by radial charge densities
$P^{(1,2,e)}(r)\equiv 4\pi r^{2}\cdot \rho^{(1,2,e)}(r).$
According to Part 3, each of the functions
$|P^{(1,2,e)}(r)|$ has at least one strongly expressed maximum,
whose position is an important characteristic of the distribution of
charge in the neighborhood of the probe particle.

In order to demonstrate the very large differences between the
alternative and DH-like characteristics, we will compare the asymptotic
behaviour of the potential $\Phi^{(1,2,e)}(r)$ and DH-like potential $\Phi^{(1,2,e)}_{DH}(r)$.
For this purpose we will take into account that from Eqs. (\ref{eq:rhoas}),
(\ref{eq:nease}), and the Supplementary material E, it follows that
\begin{equation}\label{eq:comp1}
 \Phi^{(1,2)}(r)\sim Z_{1,2}e \cdot
\frac{e^{-\kappa_{as;1,2}(r-r_{b;1,2})}}{r},\quad r>r_{b;1,2}, \quad \Phi_{DH}^{(1,2,e)}(r) \sim Z_{1,2,e}e\cdot \frac
{e^{-\kappa_{DH}r}}{r}, \quad r>0
\end{equation}
where the ion screening constants $\kappa_{as;1,2}$ are given by Eqs. (\ref{eq:kappa12as}), (\ref{eq:46}) and (\ref{eq:rhoe})
and the DH screening constant $\kappa_{DH}=(\kappa_{0;1}^{2}
+\kappa_{0;2}^{2} +\kappa_{0;e}^{2})^{1/2}$, where $\kappa_{0;1,2}$
and $\kappa_{0;e}$ are determined by Eqs. (\ref{eq:kappa12as}) and
(\ref{eq:46}).

Here, we will consider the case of classical plasma with $T_{i}= T_{e}=T$, where
$\partial\mu_{1,2,e}/\partial N_{1,2,e} = kT/N_{1,2,e}$ and,
consequently, the relations
\begin{eqnarray}\label{eq:comp5}
\frac{\kappa_{as;1,2}}{\kappa_{DH}}\equiv\frac{r_{DH}}{r_{as;1,2}}= \frac
{[Z_{1,2}^{2}N_{1,2}(1-\alpha_{e;1,2})(1-\alpha_{i})]^{\frac{1}{2}}}{(Z_{1}^{2}N_{1}
+ Z_{2}^{2}N_{2} + N_{e})^{\frac{1}{2}}},
\end{eqnarray}
are valid. These relations show that the ion asymptotic
screening constants $\kappa_{as;1,2}$ always have to be
significantly smaller than $\kappa_{DH}$, and at the same time the
corresponding screening radii $r_{as;1,2}$ always have to be
significantly larger then $r_{DH}$.
It is important that a similar result obtained in Part 2 was noted there
as an evident shortcoming of the DH solution.

From Eq. (\ref{eq:comp5}) it follows that $\Phi^{(1,2,e)}(r)$ has a completely different
asymptotic behavior compared to $\Phi_{DH}^{(1,2,e)}(r)$. Let us note that we will reach the same conclusion
by comparing the behavior of the radial charge density $P^{(1,2,e)}(r)$ to its DH-like analog.

As the main characteristics of the considered plasmas we take here the probe particle mean potential energies
$U^{(1,2,e)}$ which later get identified with the mean potential energies of ions in the real plasmas.
In order to determine these ion energies $U^{(1,2,e)}$ it is necessary to know the values
of the potential $\varphi^{(1,2,e)}$.
For the calculation of $\varphi^{(1,2,e)}$ the equations (\ref{eq:5}) are taken in the form
\begin{equation}
\label{eq:fi}
\varphi^{(1,2,e)}=\int_{0}^{\infty}\frac{\rho^{(1,2,e)}(r)}{r}4\pi r^{2}dr=4\pi\int_{0}^{\infty}\rho^{(1,2,e)}(r)dr,
\end{equation}
where the charge density $\rho^{(1,2,e)}(r)$ is described in detail in Supplementary material C. After
the determination of $\varphi^{(1,2,e)}$, the mentioned energies are obtained by means of Eq.~(\ref{eq8}).
Let us remind that the case $Z_{1}=Z_{2}=1$ can correspond to the case of plasma with the ion H$^{+}$, He$^{+}$(1s), etc.
Within this work the energies $U^{(1)}$ and $U^{(2)}$ are determined for two cases:
$Z_{1}=1$ and $Z_{2}=2$ and $Z_{1}=Z_{2}=1$. In both cases the calculations are performed
for plasmas with the electron densities $10^{16} \rm{cm^{-3}} \le N_{e} \le 10^{20} \rm{cm^{-3}}$, for $T=3\cdot 10^{4}$ K.
The obtained results are presented in Tabs 5 and 6 in the online material with quite small ion-density steps:
from $0.5 \cdot 10^{16} \rm{cm^{-3}}$ in the region of electron density $10^{17} \rm{cm^{-3}}$ to $0.5 \cdot 10^{19} \rm{cm^{-3}}$
in the region of electron density $10^{20} \rm{cm^{-3}}$.
In order to investigate the dependance of the energies of the systems on the temperature
the calculations of $U^{(1)}$ and $U^{(2)}$ were performed here in the same regions of electron densities
and with the same ion-density steps as it Tabs 5 and 6, but for the temperatures
$T=1\cdot 10^{4}$ K, $1.5\cdot 10^{4}$ K, $2\cdot 10^{4}$ K and  $2.5\cdot 10^{4}$ K. The corresponding results are presented in the tables 7-14
in the online material. From these results one can see that the potential energies $U^{(1)}$ and $U^{(2)}$ are sensitive
to considerable lowering of the temperature ($T=1\cdot 10^{4}$K).
Let as note that, for the readers' convenience, the potential energies $U^{(1)}$ and $U^{(2)}$ in all the tables are given in [eV].

Remark: at this point all calculations are also performed in the case $T_{i}=T_{e}=T$.

\subsection{Interpretation of the obtained results}

\begin{figure*}
\begin{center}
\includegraphics[width=0.49\textwidth]{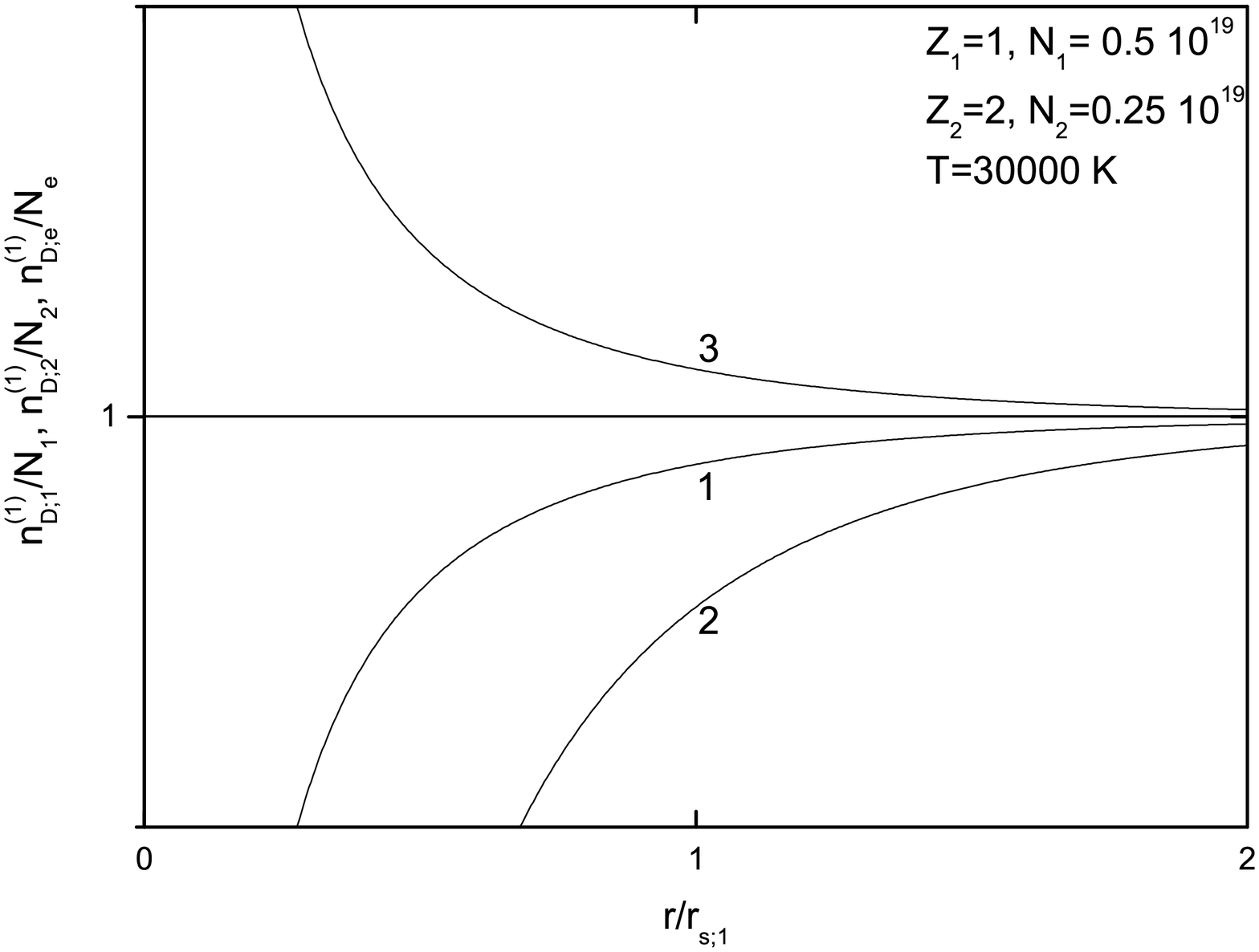}
\includegraphics[width=0.46\textwidth]{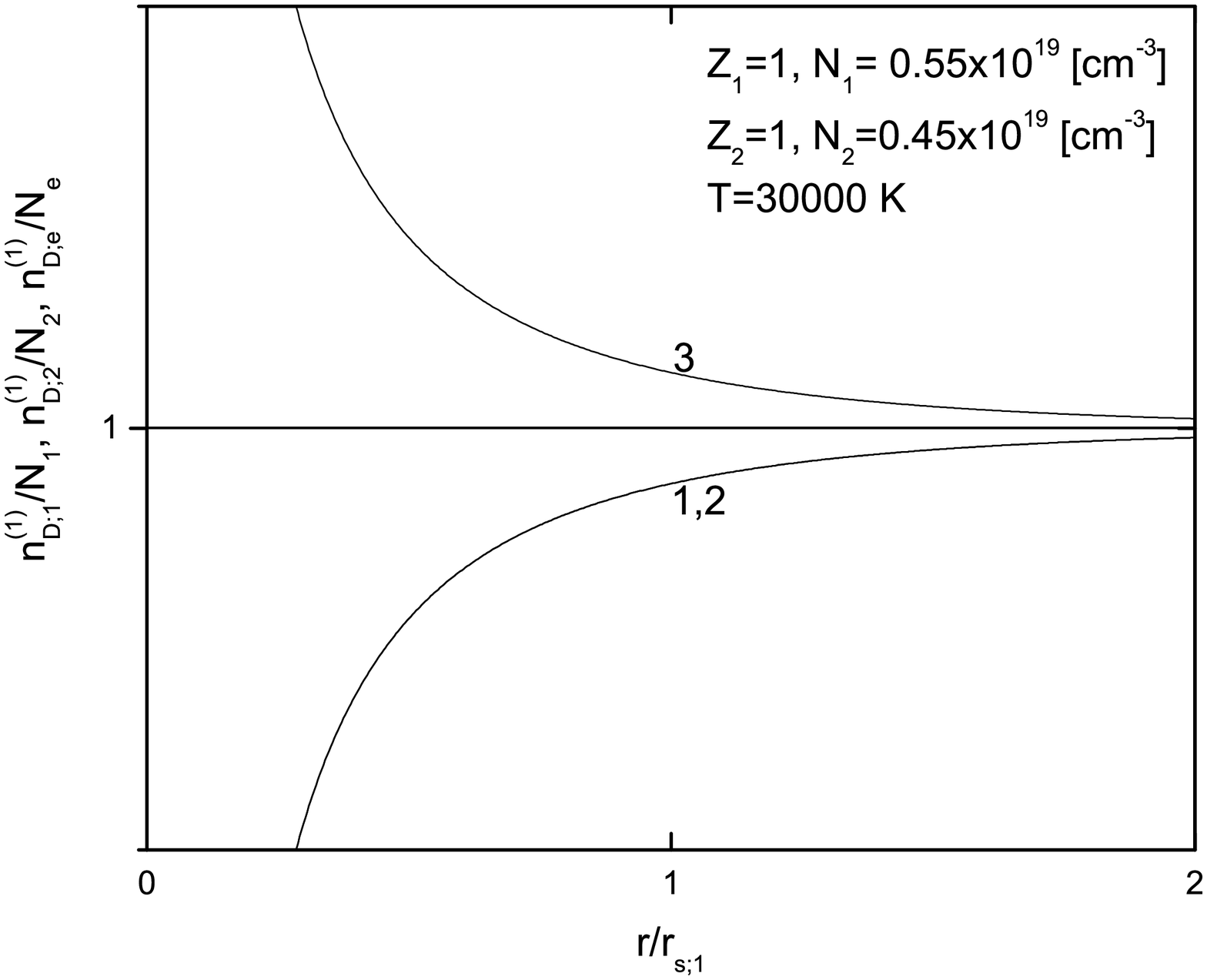}
\caption{\hspace{0.5in} a) \hspace{1.5in} b) \newline a) The reduced DH
densities $n_{D;1}^{(1)}(r)/N_{1}$ (in figure curve marked with 1), $n_{D;2}^{(1)}(r)/N_{2}$ (curve marked with 2) and
$n_{D;e}^{(1)}(r)/N_{e}$ (curve marked with 3) in the case $Z_{1}=1$, $Z_{2}=2$ and
$T_{i}=T_{e}=T$, where $T=30000K$., b) The reduced DH
densities $n_{D;1}^{(1)} (r) / N_{1}$ (in figure curve marked with 1), $n_{D;2}^{(1)} (r) / N_{2}$ (curve marked with 2),
$n_{D;e}^{(1)} (r) / N_{e}$ (curve marked with 3) in the case $Z_{1}=Z_{2}=1$ and $T_{i}=T_{e}=T$,
where $T=30000K$.}
\label{fig:nd_z}
\end{center}
\end{figure*}

Results such as those presented in the previous subsection (see \ref{eq:comp5})
might leave an impression that they mean an absolute advantage over the DH-like methods (or other similar ones).
Such an impression will be wrong since, as a matter of fact, from the presented results follows only an absolute advantage
of the presented method in the case of a system of the "closed" type.
In the same context it is necessary to interpret the phenomenon which were described in the Introduction
concerning the results presented in the Figs.~\ref{fig:htbpprof}a and \ref{fig:htbpprof}b. Namely, that phenomenon can be interpreted
as physically unacceptable when DH or DH-like method is used on the system of the "closed" type.
However the system gets treatment as "closed"-type only
in the case when it is described by means of the above mentioned (the point 2.2) additional conditions.
Consequently, in the opposite case (when additional conditions are absent) the system has to be treated as
an "open" one and can be successfully described by means of DH-like or similar methods. Hence the conclusion follows
that the way of describing the physical system depends on its treatment i.e. as an "open" type or a "closed" type.
Concerning this, see Fig.~\ref{fig:nd_z}a,b which shows the results of the application of DH method
to the considered plasma (a- $Z_1=1$ and $Z_2=2$, b- $Z_1=Z_2=1$).
It is useful to compare this figure with the Figs.~\ref{fig:htbp11} and \ref{fig:htbp113}.
In this context let us draw attention to the fact that such a treatment itself has to bee determined
based on the properties of the considered system and the physical problem which
can be solved using that system.

At the end of this point it would be useful to linger on such
influence of the additional conditions on the properties of the obtained solutions which could be treated
as manifestation of their deviation from thermodynamic equilibrium. Here we mean the necessity
of substituting the equation obtained from the condition of thermodynamic equilibrium by
the equations obtained in different ways - the electron charge density $n_{e}^{(1,2)}$ in the region
of large $r$, and the ion charge density  $n_{1,2}^{(2,1)}(r)$ - also in the region of large $r$.
However, we draw attention to the fact that all changes have purely phenomenological character
and do not influence the thermodynamic properties of the considered gases:
gas with temperature $T$ remains gas with the temperature $T$.
Apart from that, we will remind of the fact that in all regions of $r$, where the considered components
can be treated independently from one another, their state was described by means of equations
obtained from the condition of thermodynamic equilibrium.

From above presented material it follows that source basic model can generate only DH or some DH-like metod.
In this sense this model has already exhausted its potential but as one can see it enables,
with a minimal deviation from the basic model (in the area of mathematical apparatus)
to leave DH-like sphere and to develop a new model method of describing
the plasma's inner electrostatic screening.

\subsection{The possible ion-ion probe systems}
\label{ion-ion}
From the presented matter it follows that the main properties
of the considered three-component system come about from the analogy
with the properties of positron-ion probe system. This fact certainly
deserves one special additional consideration.
Concerning this it is useful to note the fact which was
established by means the molecular dynamic (MD) simulation of a
dense electron-proton plasma in \cite{rei04}. In this work
some characteristics of the considered plasma were determined as the
results of averaging over all ion configurations possible under the considered conditions,
for different values of the ratio $m_{e}/m_{p}$, where $m_{e}$ and $m_{p}$ are the
electron and proton masses. These values were changed from $1/1836$
to $1/100$, but the changes of the results of MD simulations could
be neglected. This can be very interesting even just for the
similarity of the procedures which were used here and in \cite{rei04}.
However, it can be particularly important under the
assumption that the similar conclusion is valid in the case of
plasma which contains electrons and ions of some heavy atoms,
particulary if the non-negligible
probability is taken into account that the value of the mentioned ratio can be even greater
than $1/100$. Namely, under this assumption Eq. (\ref{eq:121ion})
could be confirmed directly for such more realistic systems as
plasmas whose ion components contain H$^{+}$ and ions of atoms
heavier than Tc, for example Ag$^{2+}$, or D$^{+}$ and
ions of atoms heavier than Au, for example Hg$^{2+}$.

\section{Conclusions}
As one of the most important results of this research
a classification of physical systems (electrolytes, dusty plasmas)
is performed in this work based on consideration, or lack thereof, of a few special additional conditions. The system
considered here, as well as other systems which are described by means of the mentioned additional conditions
are treated here as the systems of the "closed" type, while the systems where
that conditions are fully neglected - as being of the "open" type. As the object of investigation here
fully ionized electron-ion plasma is chosen with positively charged ions of two different kinds,
including here the plasmas of higher non-ideality. The direct aim of this
work is to develop, within the problem of the finding of the mean potential energy of the charged particle for such plasma, a new model self-consistent method of describing of the electrostatic screening
which includes all the necessary additional conditions.
With minimal derivation from the source basic model (in the area of mathematical apparatus) this aim has been realized here.
Within the method presented such extremely significant phenomena as the electron-ion and
ion-ion correlations are included in the used model, and all types of the necessary conditions are clearly defined.
The characteristics of the considered plasmas in a wide region of the electron densities and temperatures are calculated.
All obtained results, including the comparing with the systems of the "open" type are specifically discussed
at the end of this work.
Here the case of the three-component systems was considered which is especially important since further
increase of the number of the ion components did not cause appearance of any new phenomena.

\section*{Acknowledgments}
The authors would like to express their gratitude to Professor Y. Vitel
for many years of cooperation and very helpful initial discussions,
Professor V.E. Fortov for truly priceless help in this research,
and also to Professor V.M. Adamyan,
for his assistance in the calculations in the region of very dense plasmas.
The authors are thankful to the Ministry of Education, Science and
Technological Development of the Republic of Serbia for the support
of this work within the projects 176002 and III44002.

\section*{References}


\newcommand{\noopsort}[1]{} \newcommand{\printfirst}[2]{#1}
  \newcommand{\singleletter}[1]{#1} \newcommand{\switchargs}[2]{#2#1}

\newpage
\appendix
\section{The ion densities in the regions of large and middle $r$}
\label{sec:mdmi}

\subsection{The imagined system}
\label{sec:pos}

{\bf The auxiliary expressions.}
In accordance with the the main text, an imagined positron-ion system is considered here
as an example of the necessary auxiliary system.
This example is useful since such a system can
be described practically in the same way as the electron-ion one in Part 2.

Firstly, we will consider an infinite homogenous system $S_{in}^{(+)}$ which
contains free ions and positrons with charges $Z_{i}e>0$ and $e$
respectively, and a non-structured negatively charged background.
It is assumed that the ion and positron components are treated as ideal
gases in the states of the corresponding thermodynamic equilibrium
and are characterized by the mean local
densities and temperatures $N_{i}$, $T_{i}$, $N_{+}$ and $T_{+}$, and that the local
background charge density is equal to $-e\cdot(Z_{i}N_{i}+N_{+})$, which provides
electro-neutrality of this initial system as a whole.

\begin{figure}[h!]
\begin{center}\includegraphics[width=0.8\columnwidth,
height=0.6\columnwidth]{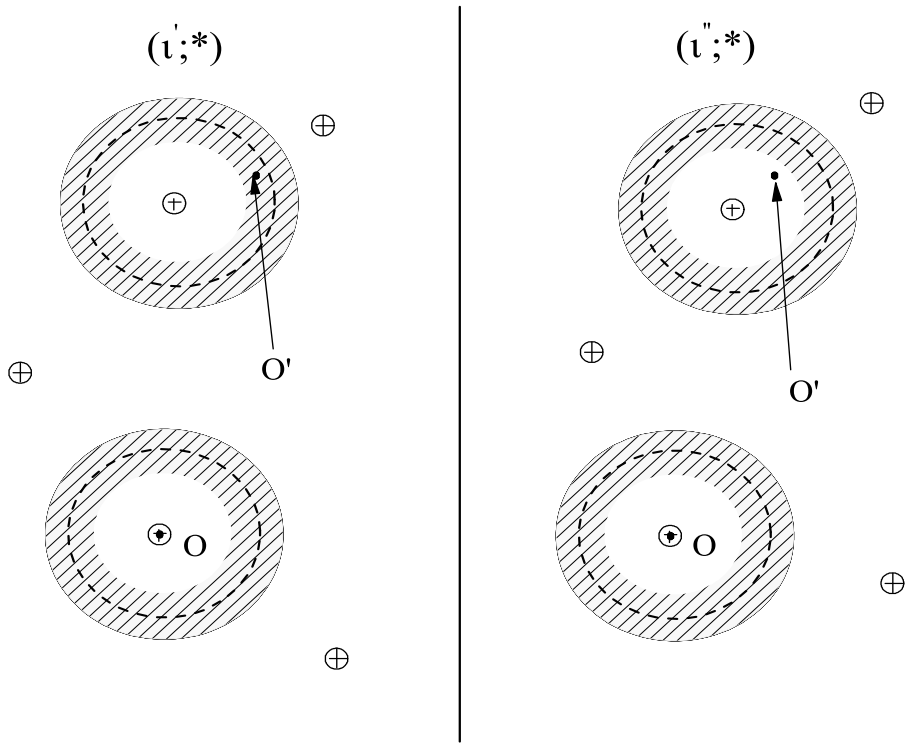}
\caption{Figure shows how the observation point $O'$ could be found
u in one of the ion configuration ($i';*$) in the region of maximum of positron density,
and in other ion configuration ($i'';*$)- in the region of minimum of positron density. The point $O$ is the coordinate origin}
\label{fig:htbpip}
\end{center}
\end{figure}

Then, we will introduce the auxiliary system $S_{a}^{(+)}$, obtained from $S_{in}^{(+)}$ by
replacing one of the ions with the probe particle, which is fixed in the origin of the
reference frame (the point $O^{+}$ in Fig. \ref{fig:htbpip}) and has a charge $Z_{p}e=Z_{i}e$.

\begin{figure}[h!]
\begin{center}\includegraphics[width=0.85\columnwidth,
height=0.6\columnwidth]{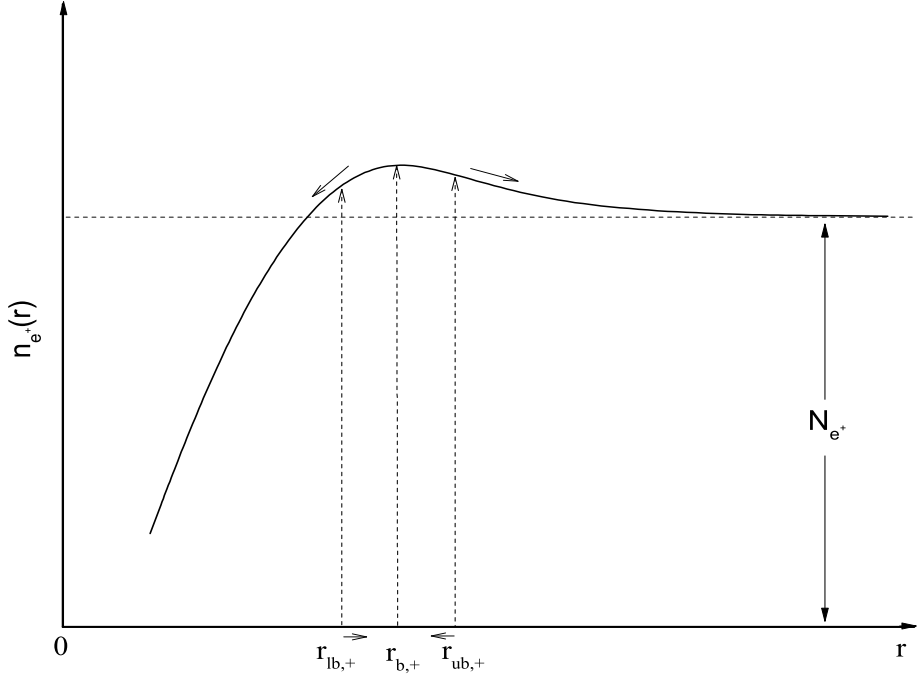}
\caption{
The expected profile of the mean local positron density in probe positron ion system.}
\label{fig:A1}
\end{center}
\end{figure}
Here we will take into account the fact that the ion component in the system $S_{a}^{(+)}$
can be treated in two ways at the same time:
as an ideal gas in relation to the used coordinate system and as a group of immobile
particles which are located in the whole space at discrete points.
All such points together are treated here as an individual ion configuration, under
the assumption that, although the positions of the heavy particles (ions) do
change with time, the distribution of the light particles (positrons)
succeeds in following these changes. Therefore in the case of an ion
configuration $(i;*)$ the system $S_{a}^{(+)}$ will be characterized by the
local ion and positron densities, $n_{i}^{*}(\vec{r})$ and $n_{+}^{(*)}(\vec{r})$,
and the positron chemical potential $\mu_{e^{+}}(n_{+}^{(*)}(\vec{r}),T_{+})$,
where $\vec{r}$ is the radius-vector of the observed point.
It is understood that $n_{i}^{*}(\vec{r})$ is equal to
zero everywhere excluding the ions and the positions, where it is equal to the corresponding
delta-functions. The density $n_{+}^{(*)}(\vec{r})$ has to be determined by means of the
equation, which is obtained from the condition of conservation of
thermodynamical equilibrium for the positron component. This condition is given by
\begin{equation}\label{eq:pozlcp}
\mu_{e^{+}}(n_{+}^{(*)}(\vec{r}),T_{+})+e\cdot \Phi^{(+;*)}(\vec{r})=
\mu_{e^{+}}(n_{+}^{(*)}(\vec{r_{st}}),T_{+})+e\cdot \Phi^{(+;*)}(\vec{r}_{st}),
\end{equation}
where the potential $\Phi^{(+;*)}(\vec{r})$ describes the electrostatic field of
the whole considered system in the case of the ion configuration $(i;*)$, and
$\vec{r}_{st}$ determines the position of some fixed (starting) point.
Consequently, the mean ion and positron densities, $n_{i}(r)$ and $n_{+}(r)$, have to be
considered as the results of the corresponding averaging of $n_{i}^{*}(\vec{r})$ and
$n_{+}^{(*)}(\vec{r})$ over all possible  ion
configurations (for the considered system).
It is assumed that they satisfy the usual boundary conditions
\begin{equation}\label{eq:bcond}
\lim\limits_{r \to \infty}n_{i}(r)=N_{i}, \qquad
\lim\limits_{r \to \infty}n_{+}(r)=N_{+},
\end{equation}
and provide electro-neutrality of the auxiliary system.

Two facts are important for the further considerations. First of them is a
consequence of the repelling character of the positron-ion and positron-probe
particle interactions, which causes that the inner parts
of the self-spheres of the probe particle ($r<r_{s;i}= Z_{i}/(Z_{i}N_{i}+N_{e^{+}})$)
and all the ions represent the regions of
the minimal local positron density ($n_{+}^{(*)}(\vec{r})<< N_{+}$), while the
very next layers (which contain the self-sphere surfaces) are the regions of the
maximal density ($n_{+}^{(*)}(\vec{r}) >N_{+}$). The other fact is a consequence of
the changes of the ion configurations, which cause that only the inner part of
the probe particle's self-sphere and its next layer (fixed layer) stay
practically always the regions of respectively the minimal and the maximal values of
$n_{+}^{(*)}(\vec{r})$. These facts, which are illustrated by
Fig.~\ref{fig:htbpip} on the examples of two different ion configurations, make it possible to
estimate the profile of $n_{+}(r)$ which is shown in Fig.~\ref{fig:A1}.
It is understood that the region of $r$ in the
neighbourhood of the point $r_{b;+}$ in Fig.~\ref{fig:A1} corresponds to
the central part of the fixed layer in Fig.~\ref{fig:htbpip}. It suggests that we
can find the points $r_{ub;+}>r_{b;+}$ and $r_{lb;+}<r_{b;+}$,
which are very close to $r_{b;+}$ and located so that $n_{+}(r)$ monotonously
decreases: with the increase of $r$ in the region $r>r_{ub;+}$, and with
the decrease of $r$ in the region $r<r_{lb;+}$. Here we will present the results of
the performed averaging of the relevant quantities (over all possible ion
configurations) separately in each of the two mentioned regions, considering that
$|\vec{r}_{st}| =r_{ub;+}$ or $r_{lb;+}$.\\

\noindent {\bf The region of large $r$.} Following Part 2, we will transform Eq.
(\ref{eq:pozlcp}) in the region $r_{ub;+} < r <\infty$ by means of the usual
linearization procedure. After the multiplication of the obtained result by
$e$, it is transformed to the equation
\begin{equation}\label{eq:pozl}
n_{+}^{(*)}(\vec{r})- n_{+}^{(*)}(\vec{r}_{st})= \frac{e}
{\partial\mu_{+}/\partial N_{+}}[\Phi^{(+;*)}(\vec{r})- \Phi^{(+;*)}
(\vec{r}_{st} )], \quad\frac {\partial\mu_{+}}{\partial N_{+}} \equiv
\left[\frac{\partial\mu_{+} (n,T_{+})}{\partial
n}\right]_{n = N_{+}}.
\end{equation}
It is understood that in this region the averaging of the left and right sides
of this equation is performed together with $n_{i}^{*}(\vec{r})$. The way of
obtaining Eq. (\ref{eq:pozl}) causes it to be applicable in the part of space
which does not contain the interior of the probe particle's or any ion's self-sphere.
Therefore we will take into account the facts that: the behaviour of
$n_{+}^{(*)}(\vec{r})$ within an ion's self-sphere has to be similar
to the that within the probe particle's sphere (since $Z_{i}e=Z_{p}e$), and
the procedure of determination of $n_{+}^{(*)}(\vec{r})$ in the vicinity of the probe
particle has to be similar to the one developed in Part 1. We will present
the result of this averaging in the form
\begin{equation}\label{eq:exam0}
n_{e^{+}}(r)-n_{e^{+}}(r_{st}) =K_{0}+ K_{1}\cdot y+
K_{2}\cdot y^{2}+..., \quad y\equiv [n_{i}(r)-N_{i}]/N_{i}.
\end{equation}
In accordance with Part 1 and 2 we can consider that $y<0$ and $|y|<<1$ for any
$r>r_{ub;+}$ which makes it possible to keep here only the first two members. Then,
due to Eq. (\ref{eq:bcond}) we have it that $K_{0}=[N_{+} - n_{+}(r_{st})]$.
In accordance with this Eq. (\ref{eq:exam0}) is transformed to the equation
\begin{equation}
n_{e^{+}}(r) - N_{+}= K_{1}\cdot [n_{i}(r) - N_{i}]/N_{i}, \qquad
r>r_{st}=r_{ub;+}.
\end{equation}
Here one of the differences between the electron-ion and
positron-ion systems manifested. Namely, in the first case the coefficient, which
corresponds to $K_{1}$, had to be grater than zero. However, the behaviour of the
positrons in the considered case is directly opposite to that of the electrons
in the electron-ion case. As a consequence, the coefficient $K_{1}$ has to
be smaller than zero. In accordance with this we will take it that $K_{1} =
-\alpha_{+}Z_{i}\cdot N_{i}$ and will present the last equation in the form
\begin{equation}\label{eq:exam1}
n_{+}(r) - N_{+}= - \alpha_{+}\cdot Z_{i}[n_{i}(r) - N_{i}],
\qquad r_{ub;+} < r < \infty,
\end{equation}
where the new unknown parameter is $\alpha_{+}>0$. This means that in the considered
region we can determine both ion and positron mean local densities using Eq.
(\ref{eq:exam1}) and two equations which are similar to Eq. (\ref{eq:121}) for
$n_{1,2}^{(1,2)}$ and Eq. (51) from Part 1 for the electrostatic potential. Let
us note that $n_{+}(r)$ increases in the region $r_{ub;+} < r < \infty$ from
$n_{+}(\infty)=N_{+}$ to $n_{+}(r_{ub;+}) > N_{+}$ with the decrease of $r$.
Consequently, this region can be treated as one where the positron-ion
correlation dominates over the influence of the presence of the fixed positively
charged probe particle, and the parameter $\alpha_{+}$ - as the coefficient of
positron-ion correlation.\\

\noindent {\bf The region of middle $r$.} In the other mentioned region, i.e.
$r_{i}<r<r_{lb;+}$, we will first perform averaging of the left and
the right sides of the condition (\ref{eq:pozlcp}), taking into account the fact that this
region corresponds to the inner part of the fixed layer in Fig. \ref{fig:A1}.
Consequently, it can be treated as the part of the region $0<r<r_{lb;+}$
where the influence of the presence of the considered probe particle dominates
over the positron-ion correlation. Therefore that we will assume that in this
region the contribution of such ion configuration that causes significant
oscillations of the $n_{+}^{(*)} (\vec{r})$ values can be neglected. As a
consequence, we can take the positron chemical potential in Eq.
(\ref{eq:pozlcp}) as
\begin{eqnarray}
\mu_{e^{+}}(n_{+}^{(*)}(\vec{r'}),T_{+})= \mu_{e^{+}}(n_{+}(r'),T_{+}) +
\frac{\partial\mu_{+}}{\partial n_{+}} \cdot
[n_{+}^{(*)}(\vec{r'})-n_{+}(r')], \nonumber \\ \frac{\partial\mu_{+}}
{\partial n_{+}} \equiv \left[\frac{\partial\mu_{+} (n,T_{+})}
{\partial n}\right]_{n = n_{+}(r')}
\end{eqnarray}
where $\vec{r'}=\vec{r}$ or $\vec{r_{st}}$. After that, as the result of
averaging Eq. (\ref{eq:pozlcp}) in the region $r_{s;i} < r < r_{lb;+}$ we
obtain the equation
\begin{equation}
\mu_{e^{+}}(n_{+}(r),T_{+})+e\cdot \Phi^{(+)}(r)=
\mu_{e^{+}}(n_{+}(r_{st}),T_{+})+e\cdot \Phi^{(+)}(r_{st}),
\quad r_{st}=r_{lb;+},
\end{equation}
where $\Phi^{(+)}(r)$ denotes the result of the averaging of
$\Phi^{(+;*)}(\vec{r})$. Then, this equation will be transformed by means of
the corresponding linearization procedure. As the result we obtain the
equation in the form
\begin{equation}\label{eq:121pos}
n_{+}(r)-n_{+}(r_{lb;+})=-\frac{e}{\partial \mu_{+}/\partial N_{+}}
\cdot\left[\Phi^{(+)}(r) - \Phi^{(+)}(r_{lb;+}) \right],
\end{equation}
where $\Phi^{(+)}(r)$ should be determined by means of Eq. (\ref{eq:6}).
Consequently, in this region $n_{+}(r)$ can be determined in the same way as
$n_{i}(r)$, i.e. by means of an equation similar to Eq. (\ref{eq:121}).

Within this procedure the equations for $n_{+}(r)$ in the final form are
obtained after the extrapolations of Eqs. (\ref{eq:exam1}) and (\ref{eq:121pos})
to the point $r_{b;+}$ in the middle of the transition layer $r_{lb;+} < r <
r_{ub;+}$. This means that the already obtained equations become final ones by
replacing $r_{lb;+}$ and $r_{ub;+}$ with $r_{b;+}$. One can see that the
solution of Eqs. (\ref{eq:exam1}) and (\ref{eq:121pos}), with $r_{lb;+}=r_{ub;+}
=r_{b;+}$, makes possible satisfying a condition similar to Eq.
(\ref{eq:three}).

\subsection{The considered systems}
\label{sec:ion}

The equations, which are needed for the systems $S_{a}^{(1,2)}$, are obtained
here by replacing the origin designations in Eqs. (\ref{eq:exam1}) and
(\ref{eq:121pos}) with the ones corresponding to
the ion densities $n_{2}^{(1)}$ and $n_{1}^{(2)}$ (i.e. $e$ with $Z_{2,1}e$ etc.).
In such a way Eqs.  (\ref{eq:exam1}) and (\ref{eq:121pos}) are transformed to the equations
\begin{eqnarray}\label{eq:121ion}
n_{2,1}^{(1,2)}(r) - n_{2,1}^{(1,2)}(r_{b;1,2}) = -
\frac{Z_{2,1}e}{\partial \mu_{2,1}/\partial N_{2,1}}
\cdot\left[\Phi^{(1,2)}(r) - \Phi^{(1,2)}(r_{n;1,2}) \right],
\nonumber \\ l_{s;1,2}<r<r_{b;1,2},
\end{eqnarray}
\begin{equation}\label{eq:dif}
Z_{2,1}[n_{2,1}^{(1,2)}(r)-N_{2,1}]=-\alpha_{i}\cdot
Z_{1,2}[n_{1,2}^{(1,2)}(r)-N_{1,2}],
\qquad r_{b;1,2} < r < \infty,
\end{equation}
where $\Phi^{(1,2)}(r)$ is given by Eq. (\ref{eq:6}), and $\alpha_{i}>0$ has the
meaning of the coefficient of the ion-ion correlation. Let us note that within a
simple model method (which is developed here) only the procedure used - the
extrapolation of the obtained equations to some middle point - provides the
self-consistence of the final expressions for the ion densities $n_{2}^{(1)}$
and $n_{1}^{(2)}$.

The argumentation for accepting the Eq. (\ref{eq:121ion})
is based on the qualitative similarity of the behavior of the positively charged
components in the systems $S_{a}^{(1,2)}$ and $S_{a}^{(+)}$, as well as on the
facts that the form of the presented equations is determined by the parameters
$r_{b;1,2}$ and $\alpha_{i}$, and that the solutions of these equations provide
the satisfaction of Eq. (\ref{eq:three}). However, in the
considered case we have also two parameters, $r_{b;1,2}$ and $\alpha_{i}>0$,
which can be strongly determined by independent conditions; we mean
Eq. (\ref{eq:three}) for $r_{b;1,2}$, and Eq. (\ref{eq:addc}) for $\alpha_{i}$.
Therefore
just these equations are accepted here as the ones necessary for
describing all the ion components in the systems $S_{a}^{(1,2)}$.

\section{The charge densities in the regions of middle $r$}
\label{sec:mdmS}

Taking that $\rho^{(1,2)}(r) = s^{(1,2)}(r)/r$ in the regions $l_{s;1,2} < r <
r_{b;1,2})$, we will transform Eq. (\ref{eq:kappai1}) to the integral equation
\begin{equation}\label{eq:intS}
s^{(1,2)}(r) = \kappa_{i}^2 \cdot\left[\int\limits_{r}^{r_{b;1,2}}
{s^{(1,2)} (r')(r' - r)}dr' + \rho_{1,2}\cdot\frac{1 -
\kappa_{as;1,2}r_{b;1,2}} {\kappa_{as;1,2}^2}\cdot(r_{b;1,2} -
r)\right] + r\cdot\rho_{1,2},
\end{equation}
where $\rho_{1,2}\equiv \lim\limits_{r \to r_{b;1,2}} [s^{(1,2)}(r)/r]$.
Then, performing on it by the operator $d^2/dr^2$, we obtained the needed
differential equation, namely: $d^2s^{(1,2)}(r) /dr^2 = \kappa_{i;1,2}^2 \cdot
s^{(1,2)}(r)$. As it is known, its solution is given by the relation:
$s^{(1,2)}(r) = c'_{1,2}\cdot e^{-\kappa_{i;1,2}r} + c"_{1,2}\cdot
e^{\kappa_{i;1,2}r}$. With such $s^{(1,2)}(r)$ from  Eqs. (\ref{eq:intS}) it is
obtained the equation
\begin{equation}\label{eq:AB}
r\cdot q_{1,2}(c',c") = r_{b;1,2}\cdot q_{1,2}(c',c"),
\end{equation}
where: $q_{1,2}(c',c") \equiv (c'_{1,2} \cdot e^{-\kappa_{i}r_{b;1,2}} + c"_{1,2}
\cdot e^{\kappa_{i}r_{b;1,2}}) \cdot [\kappa_{i}^2/\kappa_{as;1,2}^2 +
(\kappa_{i}/\kappa_{as;1,2}) \cdot \kappa_{i}r_{b;1,2} - 1 ] -(c'_{1,2} \cdot
e^{-\kappa_{i}r_{b;1,2}} \cdot \kappa_{i}r_{b;1,2} -c"_{1,2} \cdot
e^{\kappa_{i}r_{b;1,2}} \cdot \kappa_{i}r_{b;1,2})$. Since this equation has to
be valid not only for $r=r_{b;1,2}$, but also for $r < r_{b;1,2}$, it has the
sense only if $q_{1,2}(c',c")\equiv 0$, i.e. when: $c"_{1,2} = - c'_{1,2} \cdot
e^{-2\kappa_{i;1,2}r_{b;1,2}}\cdot f_{1,2}$. Consequently, we have that
\begin{equation}\label{eq:relAB}
s^{(1,2)}(r) = c'_{1,2}\cdot e^{-\kappa_{i;1,2}r_{b;1,2}}\cdot
[e^{\kappa_{i;1,2}(r_{b;1,2}-r)} - f_{1,2} \cdot
e^{-\kappa_{i;1,2}(r_{b;1,2}-r)}],
\end{equation}
where $f_{1,2}$ is given in subsection \ref{sec:i12a} of Section \ref{sec:i12} by
Eq. (\ref{eq:ABCD}). From here it follows the expression (\ref{eq:rhorho}) for the
solution of Eq. (\ref{eq:kappai1}), where $c'_{1,2}$ is presented in an adequate
form.

\section{The charge densities in the cases $(i1)$, $(i2)$ and $(e)$ }
\label{CD}

On the bases of the matter from the main text it follows that $\rho^{(1,2)}(r)$ can be presented in the whole
space in the form, under the condition $Z_{1}\le Z_{2}$, which guaranties that
$r_{0;1}^{(1,2)}\le r_{0;2}^{(1,2)}$
\begin{equation}\label{eqrhofin}
\rho^{(1,2,e)}(r) = \rho_{0}^{(1,2,e)}(r) + \rho_{s}^{(1,2,e)}(r), \
\end{equation}
where the both members depends of the considered case.

I the symmetrical case, $Z_{1} = Z_{2}$, here it is use the calculation
scheme where the first member in the case $(i1)$ is given by the expressions:
\begin{equation}\label{eqmemb11}
\rho_{0}^{(1)}(r) = -e N_{e}(1 - \alpha_{e}), \qquad 0 < r < r_{1}^{(1)};
\end{equation}
\begin{equation}
\rho_{0}^{(1)}(r) =
-e\cdot\left[N_{e}-Z_{1}^{(1)}N_{1}^{(1)} + Z_{1}N_{1}\cdot A_{1}+
Z_{1}N_{1}B_{1}d_{1}r_{b;1}\cdot \frac{e^{\kappa_{i}(r_{b;1}-r)}-
f_{1}e^{-\kappa_{i}(r_{b;1}-r)}}{r}\right]
\cdot (1 - \alpha_{e}),
\end{equation}
where $r_{1}^{(1)} < r < r_{2}^{(1)}$,
\begin{equation}
\rho_{0}^{(1)}(r) = - e\cdot\left\{\begin{array}{*{20}c}
\displaystyle{
Z_{1}N_{1}B_{1}r_{b;1}\cdot \frac{e^{\kappa_{i}(r_{b;1}-r)}-
f_{1}e^{-\kappa_{i}(r_{b;1}-r)}}{r}\cdot (1 - \alpha_{e}), }
\hfill & r_{2}^{()} < r \le r_{b;1} , \hfill \\
\displaystyle{Z_{1}N_{1}C_{1}r_{b;1}\cdot \frac{e^{-\kappa_{as;1}(r-r_{b;1})}}
{r}\cdot(1 -\alpha_{i})\cdot(1 -\alpha_{e})}, \hfill & r_{b;1} < r < \infty ,
\end{array}
\right.
\end{equation}
and in the case $(i2)$ - by the similar expressions:
\begin{equation}\label{eqmemb12}
\rho_{0}^{(2)}(r) = -e N_{e}(1 - \alpha_{e}), \qquad 0 < r < r_{1}^{(2)};
\end{equation}
\begin{equation}
\rho_{0}^{(2)}(r) =
-e\cdot\left[N_{e}-Z_{1}^{(1)}N_{1}^{(1)} - Z_{2}N_{2}\cdot A_{2}+
Z_{2}N_{2}B_{2}d_{1}r_{b;2}\cdot \frac{e^{\kappa_{i}(r_{b;2}-r)}-
f_{1}e^{-\kappa_{i}(r_{b;2}-r)}}{r}\right]
\cdot (1 - \alpha_{e}),
\end{equation}
where $r_{1}^{(2)} < r < r_{2}^{(2)}$,
\begin{equation}
\rho_{0}^{(2)}(r) = - e\cdot\left\{ \begin{array}{*{20}c}
\displaystyle{
Z_{2}N_{2}B_{2}r_{b;2}\cdot \frac{e^{\kappa_{i}(r_{b;2}-r)}-
f_{2}e^{-\kappa_{i}(r_{b;2}-r)}}{r}\cdot (1 - \alpha_{e}), }
\hfill & r_{2}^{()} < r \le r_{b;2} , \hfill \\
\displaystyle{Z_{2}N_{2}C_{2}r_{b;2}\cdot \frac{e^{-\kappa_{as;2}(r-r_{b;2})}}
{r}\cdot(1 -\alpha_{i})\cdot(1 -\alpha_{e})}, \hfill & r_{b;2} < r < \infty ,
\end{array}
\right.
\end{equation}
where
\begin{equation}
d_{1,2}=\frac{\kappa_{0;1,2}^{2}}{\kappa_{0;1}^{2}+\kappa_{0;2}^{2}},
\end{equation}

\begin{equation}
\label{eq83a}
\kappa_{i}=
[(\kappa_{0;1}^{2}+\kappa_{0;2}^{2})\cdot (1-\alpha_{e})]^{1/2},
\end{equation}
while the second member in the cases $(i1)$ and $(i2)$ is given by
\begin{equation}\label{eqmembe}
\rho_{s}^{(1,2)} = -e\cdot[n_{e;s}^{(1,2)}(r)-N_{e}\cdot (1-\alpha_{e})].
\qquad 0 < r < l_{s;1,2}.
\end{equation}

In the case $(i1)$ this scheme understood that $R_{1}^{(1)}=r_{0;1}^{(1)}$
and $R_{2}^{(1)}=r_{0;2}^{(1)}$ if $r_{0;2}^{(1)} > r_{0;1}^{(1)}$.
In the opposite case $r_{0;2}^{(1)} < r_{0;1}^{(1)}$ it is taken that
$R_{1}^{(1)}=r_{0;2}^{(1)}$ and $R_{2}^{(1)}=r_{0;1}^{(1)}$.

Similarly in the case $(i2)$ we take that
$R_{1}^{(2)}=r_{0;1}^{(2)}$ and $R_{2}^{(2)}=r_{0;2}^{(2)}$,
if $r_{0;2}^{(2)} > r_{0;1}^{(2)}$.

Otherwise, if $r_{0;2}^{(2)} < r_{0;1}^{(2)}$ it is taken that
$R_{1}^{(2)}=r_{0;2}^{(2)}$ and $R_{2}^{(2)}=r_{0;1}^{(2)}$.

Finally, we have that in the case $(e)$ the first and second members in \ref{eqrhofin} are
given by the expressions
\begin{equation}
\rho_{0}^{(e)}(r) = e (1 - \alpha_{e})\cdot N_{e}, \qquad 0 < r < r_{0;e},
\end{equation}
\begin{equation}\label{eq83}
\rho_{0}^{(e)}(r) = e (1 - \alpha_{e})
\cdot \left[ N_{e} - n_{e}^{(e)}(r)\right], \qquad r_{0;e} < r < \infty,
\end{equation}
\begin{equation}
\rho_{s}^{(e)}(r) = e\cdot\left\{ \begin{array}{*{20}c}
\displaystyle{[Z_{1}n_{s;1}^{(e)}(r)+Z_{2}n_{s;2}^{(e)}(r)]
- e\cdot N_{e}\cdot(1 - \alpha_{e}), }\hfill & 0 < r \le l_{e;2} ,
\hfill \\
\displaystyle{[Z_{1}n_{s;1}^{(e)}(r)]
- e\cdot N_{e}\cdot p_{1}\cdot(1 - \alpha_{e})}, \hfill & l_{e;2} < r \le l_{e;1} ,
\end{array}
\right.
\end{equation}
where $n_{e;s}^{(1,2)}(r)$ is defined by Eqs. (\ref{eq:45}) - (\ref{eq:46}).

\section{Additional online material}
Additional online material (tables) may be found in the online version of this article.

\section{DH solutions}

Within DH method the Poisson's equation (\ref{eq5}) is used in usual way for the obtaining of the corresponding Helmholtz's
equation for the mean electrostatic potential: $\nabla^{2}\Phi_{DH}^{(1,2,e)}(r)=\kappa_{D}^{2}\Phi_{DH}^{(1,2,e)}(r)$, which applies in
the whole region $0 < r < \infty$ with the boundary conditions (\ref{eq:6}) and (\ref{eq:7}). The DH solutions of this equation,
i.e. $\Phi_{DH}^{(1,2,e)}(r)$, and DH screening constant $\kappa_{D}$ are given by relations
\begin{equation}
\label{DHD}
\Phi_{DH}^{(1,2,e)}(r)=\frac{Z_{1,2,e}e}{r}\cdot \frac{\exp{(-\kappa_{D}r)}}{r}, \quad
\kappa_{D} \equiv \frac{1}{r_{D}}=(\kappa_{0;1}^{2}+\kappa_{0;2}^{2}+\kappa_{0;e}^{2})^{1/2}.
\end{equation}
From here and Eqs. (\ref{eq:7}) and (\ref{eq8}) it follows that DH potential
$\varphi_{D}^{(1,2,e)}=-Z_{1,2,e}e \cdot \kappa_{D}$ and consequently,
DH potential energy
\begin{equation}
\label{eqA6}
U_{D}^{(1,2,e)}=-Z_{1,2,e}e \cdot \kappa_{D}=-\frac{( Z_{1,2,e}e)^{2}}{r_{D}}
\end{equation}

The solutions for DH charge and particle densities, $n_{D;1}^{(1,2,e)}(r)$, $n_{D;2}^{(1,2,e)}(r)$ and $n_{D;e}^{(1,2,e)}(r)$
are obtained by means of Eq. (\ref{DHD}).
They are given by
\begin{equation}
\label{eqA4}
n_{D;1}^{(1)}= N_1 - \frac{\kappa_{0;1}^{2}}{4 \pi} \frac{\exp{(-\kappa_{D}r)}}{r}, \quad
n_{D;e}^{(e)}= N_e - \frac{\kappa_{0;e}^{2}}{4 \pi } \frac{\exp{(-\kappa_{D}r)}}{r},
\end{equation}
\begin{equation}
\label{eqA5}
n_{D;e}^{(i)}= N_e + \frac{Z_{i}\kappa_{0;e}^{2}}{4 \pi} \frac{\exp{(-\kappa_{D}r)}}{r}, \quad
n_{D;i}^{(e)}= N_i + \frac{\kappa_{0;i}^{2}}{4 \pi Z_{i}} \frac{\exp{(-\kappa_{D}r)}}{r},
\end{equation}
and are illustrated by the figure \ref{fig:nd_z}a,b.
\begin{verbatim}
Table 1: The characteristic length $l_{s;1,2}$ in [cm] and the nondimensional electron-ion correlation coefficients $\alpha_{e,1,2}$
for the case $Z_{1}=1$ and $Z_{2}=2$ at $T=3 \cdot 10^{4}$ K $ in the region of electron densities
$10^{16} cm^{-3} \le N_e \le 10^{20} cm^{-3}; N_{1,2} in [cm^{-3}]

N_1	N_2	l_{s;1}		l_{s;2}	   \alpha_{e,1}	\alpha_{e,2}
====================================================================
5.00E15	4.75E16	1.45869E-6	1.67768E-6	0.0124	0.01959
1.00E16	4.50E16	1.45323E-6	1.67138E-6	0.0124	0.01959
1.50E16	4.25E16	1.44768E-6	1.66497E-6	0.0124	0.01959
2.00E16	4.00E16	1.44203E-6	1.65847E-6	0.0124	0.01959
2.50E16	3.75E16	1.43630E-6	1.65186E-6	0.0124	0.01959
3.00E16	3.50E16	1.43048E-6	1.64515E-6	0.0124	0.01959
3.50E16	3.25E16	1.42455E-6	1.63831E-6	0.0124	0.01959
4.00E16	3.00E16	1.41852E-6	1.63136E-6	0.0124	0.01959
4.50E16	2.75E16	1.41239E-6	1.62429E-6	0.0124	0.01959
5.00E16	2.50E16	1.40615E-6	1.61710E-6	0.0124	0.01959
5.50E16	2.25E16	1.39979E-6	1.60977E-6	0.0124	0.01959
6.00E16	2.00E16	1.39331E-6	1.60230E-6	0.0124	0.01959
6.50E16	1.75E16	1.38670E-6	1.59469E-6	0.0124	0.01959
7.00E16	1.50E16	1.37997E-6	1.58694E-6	0.0124	0.01959
7.50E16	1.25E16	1.37311E-6	1.57902E-6	0.0124	0.01959
8.00E16	1.00E16	1.36610E-6	1.57094E-6	0.0124	0.01959
8.50E16	7.50E15	1.35894E-6	1.56270E-6	0.0124	0.01959
9.00E16	5.00E15	1.35163E-6	1.55427E-6	0.0124	0.01959
9.50E16	2.50E15	1.34415E-6	1.54565E-6	0.0124	0.01959

5.00E16	4.75E17	6.76153E-7	7.78746E-7	0.03293	0.05162
1.00E17	4.50E17	6.73657E-7	7.75856E-7	0.03265	0.05119
1.50E17	4.25E17	6.71122E-7	7.72920E-7	0.03237	0.05075
2.00E17	4.00E17	6.68546E-7	7.69939E-7	0.03208	0.05031
2.50E17	3.75E17	6.65928E-7	7.66909E-7	0.03179	0.04985
3.00E17	3.50E17	6.63267E-7	7.63830E-7	0.03149	0.04939
3.50E17	3.25E17	6.60561E-7	7.60699E-7	0.03118	0.04891
4.00E17	3.00E17	6.57808E-7	7.57510E-7	0.03087	0.04843
4.50E17	2.75E17	6.55006E-7	7.54269E-7	0.03055	0.04794
5.00E17	2.50E17	6.52155E-7	7.50969E-7	0.03022	0.04743
5.50E17	2.25E17	6.49251E-7	7.47609E-7	0.02989	0.04692
6.00E17	2.00E17	6.46292E-7	7.44186E-7	0.02955	0.04639
6.50E17	1.75E17	6.43274E-7	7.40698E-7	0.0292	0.04584
7.00E17	1.50E17	6.40201E-7	7.37141E-7	0.02884	0.04529
7.50E17	1.25E17	6.37065E-7	7.33513E-7	0.02847	0.04472
8.00E17	1.00E17	6.33864E-7	7.29811E-7	0.0281	0.04413
8.50E17	7.50E16	6.30596E-7	7.26031E-7	0.02771	0.04353
9.00E17	5.00E16	6.27256E-7	7.22169E-7	0.02731	0.04290
9.50E17	2.50E16	6.23842E-7	7.18222E-7	0.02689	0.04226

5.00E17	4.75E18	3.12965E-7	3.61497E-7	0.05592	0.08690
1.00E18	4.50E18	3.11846E-7	3.60191E-7	0.05592	0.08690
1.50E18	4.25E18	3.10709E-7	3.58864E-7	0.05592	0.08690
2.00E18	4.00E18	3.09554E-7	3.57516E-7	0.05592	0.08690
2.50E18	3.75E18	3.08380E-7	3.56146E-7	0.05592	0.08690
3.00E18	3.50E18	3.07186E-7	3.54754E-7	0.05592	0.08690
3.50E18	3.25E18	3.05972E-7	3.53339E-7	0.05592	0.08690
4.00E18	3.00E18	3.04738E-7	3.51898E-7	0.05592	0.08690
4.50E18	2.75E18	3.03481E-7	3.50432E-7	0.05592	0.08690
5.00E18	2.50E18	3.02202E-7	3.48941E-7	0.05592	0.08690
5.50E18	2.25E18	3.00899E-7	3.47422E-7	0.05592	0.08690
6.00E18	2.00E18	2.99570E-7	3.45875E-7	0.05592	0.08690
6.50E18	1.75E18	2.98218E-7	3.44298E-7	0.05592	0.08690
7.00E18	1.50E18	2.96840E-7	3.42691E-7	0.05592	0.08690
7.50E18	1.25E18	2.95434E-7	3.41052E-7	0.05592	0.08690
8.00E18	1.00E18	2.93998E-7	3.39379E-7	0.05592	0.08690
8.50E18	7.50E17	2.92533E-7	3.37672E-7	0.05592	0.08690
9.00E18	5.00E17	2.91036E-7	3.35928E-7	0.05592	0.08690
9.50E18	2.50E17	2.89506E-7	3.34146E-7	0.05592	0.08690

5.00E18	4.75E19	1.44456E-7	1.67825E-7	0.11559	0.17559
1.00E19	4.50E19	1.43973E-7	1.67253E-7	0.11559	0.17559
1.50E19	4.25E19	1.43482E-7	1.66671E-7	0.11559	0.17559
2.00E19	4.00E19	1.42983E-7	1.66080E-7	0.11559	0.17559
2.50E19	3.75E19	1.42475E-7	1.65479E-7	0.11559	0.17559
3.00E19	3.50E19	1.41960E-7	1.64869E-7	0.11559	0.17559
3.50E19	3.25E19	1.41435E-7	1.64248E-7	0.11559	0.17559
4.00E19	3.00E19	1.40901E-7	1.63617E-7	0.11559	0.17559
4.50E19	2.75E19	1.40358E-7	1.62974E-7	0.11559	0.17559
5.00E19	2.50E19	1.39805E-7	1.62319E-7	0.11559	0.17559
5.50E19	2.25E19	1.39242E-7	1.61654E-7	0.11559	0.17559
6.00E19	2.00E19	1.38669E-7	1.60975E-7	0.11559	0.17559
6.50E19	1.75E19	1.38085E-7	1.60284E-7	0.11559	0.17559
7.00E19	1.50E19	1.37489E-7	1.59580E-7	0.11559	0.17559
7.50E19	1.25E19	1.36882E-7	1.58862E-7	0.11559	0.17559
8.00E19	1.00E19	1.36262E-7	1.58130E-7	0.11559	0.17559
8.50E19	7.50E18	1.35630E-7	1.57383E-7	0.11559	0.17559
9.00E19	5.00E18	1.34984E-7	1.56620E-7	0.11559	0.17559
9.50E19	2.50E18	1.34325E-7	1.55841E-7	0.11559	0.17559
=======


Table 2: The same as in table 1 but for the case $Z_{1}=Z_{2}=1$.

N_1	N_2	l_{s;1}		l_{s;2}	   \alpha_{e,1}	\alpha_{e,2}
====================================================================
5.00E15	9.50E16	1.3365E-6	1.3365E-6	0.0124	0.0124
1.00E16	9.00E16	1.3365E-6	1.3365E-6	0.0124	0.0124
1.50E16	8.50E16	1.3365E-6	1.3365E-6	0.0124	0.0124
2.00E16	8.00E16	1.3365E-6	1.3365E-6	0.0124	0.0124
2.50E16	7.50E16	1.3365E-6	1.3365E-6	0.0124	0.0124
3.00E16	7.00E16	1.3365E-6	1.3365E-6	0.0124	0.0124
3.50E16	6.50E16	1.3365E-6	1.3365E-6	0.0124	0.0124
4.00E16	6.00E16	1.3365E-6	1.3365E-6	0.0124	0.0124
4.50E16	5.50E16	1.3365E-6	1.3365E-6	0.0124	0.0124
5.00E16	5.00E16	1.3365E-6	1.3365E-6	0.0124	0.0124
5.50E16	4.50E16	1.3365E-6	1.3365E-6	0.0124	0.0124
6.00E16	4.00E16	1.3365E-6	1.3365E-6	0.0124	0.0124
6.50E16	3.50E16	1.3365E-6	1.3365E-6	0.0124	0.0124
7.00E16	3.00E16	1.3365E-6	1.3365E-6	0.0124	0.0124
7.50E16	2.50E16	1.3365E-6	1.3365E-6	0.0124	0.0124
8.00E16	2.00E16	1.3365E-6	1.3365E-6	0.0124	0.0124
8.50E16	1.50E16	1.3365E-6	1.3365E-6	0.0124	0.0124
9.00E16	1.00E16	1.3365E-6	1.3365E-6	0.0124	0.0124
9.50E16	5.00E15	1.3365E-6	1.3365E-6	0.0124	0.0124

5.00E16	9.50E17	6.2035E-7	6.2035E-7	0.02607	0.02607
1.00E17	9.00E17	6.2035E-7	6.2035E-7	0.02566	0.02566
1.50E17	8.50E17	6.2035E-7	6.2035E-7	0.02524	0.02524
2.00E17	8.00E17	6.2035E-7	6.2035E-7	0.0248	0.02480
2.50E17	7.50E17	6.2035E-7	6.2035E-7	0.02435	0.02435
3.00E17	7.00E17	6.2035E-7	6.2035E-7	0.02387	0.02387
3.50E17	6.50E17	6.2035E-7	6.2035E-7	0.02338	0.02338
4.00E17	6.00E17	6.2035E-7	6.2035E-7	0.02286	0.02286
4.50E17	5.50E17	6.2035E-7	6.2035E-7	0.02232	0.02232
5.00E17	5.00E17	6.2035E-7	6.2035E-7	0.02175	0.02175
5.50E17	4.50E17	6.2035E-7	6.2035E-7	0.02115	0.02115
6.00E17	4.00E17	6.2035E-7	6.2035E-7	0.02051	0.02051
6.50E17	3.50E17	6.2035E-7	6.2035E-7	0.01983	0.01983
7.00E17	3.00E17	6.2035E-7	6.2035E-7	0.01909	0.01909
7.50E17	2.50E17	6.2035E-7	6.2035E-7	0.0183	0.01830
8.00E17	2.00E17	6.2035E-7	6.2035E-7	0.01742	0.01742
8.50E17	1.50E17	6.2035E-7	6.2035E-7	0.01644	0.01644
9.00E17	1.00E17	6.2035E-7	6.2035E-7	0.01533	0.01533
9.50E17	5.00E16	6.2035E-7	6.2035E-7	0.01402	0.01402

5.00E17	9.50E18	2.87941E-7	2.87941E-7	0.05592	0.05592
1.00E18	9.00E18	2.87941E-7	2.87941E-7	0.05592	0.05592
1.50E18	8.50E18	2.87941E-7	2.87941E-7	0.05592	0.05592
2.00E18	8.00E18	2.87941E-7	2.87941E-7	0.05592	0.05592
2.50E18	7.50E18	2.87941E-7	2.87941E-7	0.05592	0.05592
3.00E18	7.00E18	2.87941E-7	2.87941E-7	0.05592	0.05592
3.50E18	6.50E18	2.87941E-7	2.87941E-7	0.05592	0.05592
4.00E18	6.00E18	2.87941E-7	2.87941E-7	0.05592	0.05592
4.50E18	5.50E18	2.87941E-7	2.87941E-7	0.05592	0.05592
5.00E18	5.00E18	2.87941E-7	2.87941E-7	0.05592	0.05592
5.50E18	4.50E18	2.87941E-7	2.87941E-7	0.05592	0.05592
6.00E18	4.00E18	2.87941E-7	2.87941E-7	0.05592	0.05592
6.50E18	3.50E18	2.87941E-7	2.87941E-7	0.05592	0.05592
7.00E18	3.00E18	2.87941E-7	2.87941E-7	0.05592	0.05592
7.50E18	2.50E18	2.87941E-7	2.87941E-7	0.05592	0.05592
8.00E18	2.00E18	2.87941E-7	2.87941E-7	0.05592	0.05592
8.50E18	1.50E18	2.87941E-7	2.87941E-7	0.05592	0.05592
9.00E18	1.00E18	2.87941E-7	2.87941E-7	0.05592	0.05592
9.50E18	5.00E17	2.87941E-7	2.87941E-7	0.05592	0.05592

5.00E18	9.50E19	1.3365E-7	1.3365E-7	0.11559	0.11559
1.00E19	9.00E19	1.3365E-7	1.3365E-7	0.11559	0.11559
1.50E19	8.50E19	1.3365E-7	1.3365E-7	0.11559	0.11559
2.00E19	8.00E19	1.3365E-7	1.3365E-7	0.11559	0.11559
2.50E19	7.50E19	1.3365E-7	1.3365E-7	0.11559	0.11559
3.00E19	7.00E19	1.3365E-7	1.3365E-7	0.11559	0.11559
3.50E19	6.50E19	1.3365E-7	1.3365E-7	0.11559	0.11559
4.00E19	6.00E19	1.3365E-7	1.3365E-7	0.11559	0.11559
4.50E19	5.50E19	1.3365E-7	1.3365E-7	0.11559	0.11559
5.00E19	5.00E19	1.3365E-7	1.3365E-7	0.11559	0.11559
5.50E19	4.50E19	1.3365E-7	1.3365E-7	0.11559	0.11559
6.00E19	4.00E19	1.3365E-7	1.3365E-7	0.11559	0.11559
6.50E19	3.50E19	1.3365E-7	1.3365E-7	0.11559	0.11559
7.00E19	3.00E19	1.3365E-7	1.3365E-7	0.11559	0.11559
7.50E19	2.50E19	1.3365E-7	1.3365E-7	0.11559	0.11559
8.00E19	2.00E19	1.3365E-7	1.3365E-7	0.11559	0.11559
8.50E19	1.50E19	1.3365E-7	1.3365E-7	0.11559	0.11559
9.00E19	1.00E19	1.3365E-7	1.3365E-7	0.11559	0.11559
9.50E19	5.00E18	1.3365E-7	1.3365E-7	0.11559	0.11559
============

Table 3: The main characteristic length $r_{b;1,2}$ in [cm] and nondimensional ion-ion correlation coefficients
$\alpha_{i}$ for the case $Z_{1}=1$ and $Z_{2}=2$ at $T=3 \cdot 10^{4}$ K $
in the region of electron densities  $10^{16} cm^{-3} \le N_e \le 10^{20} cm^{-3}; N_{1,2} in [cm^{-3}]


N_1	N_2	r_{b;1}	        r_{b;2}	      \alpha_{i}
====================================================
5.00E15	4.75E16	2.23180E-6	9.07625E-6	0.10
1.00E16	4.50E16	2.29610E-6	7.33734E-6	0.10
1.50E16	4.25E16	2.22942E-6	6.14375E-6	0.09
2.00E16	4.00E16	2.30726E-6	5.55587E-6	0.09
2.50E16	3.75E16	2.36990E-6	5.12077E-6	0.09
3.00E16	3.50E16	2.46042E-6	4.78738E-6	0.09
3.50E16	3.25E16	2.54994E-6	4.50535E-6	0.09
4.00E16	3.00E16	2.65264E-6	4.29048E-6	0.09
4.50E16	2.75E16	2.75416E-6	4.09322E-6	0.09
5.00E16	2.50E16	3.05134E-6	4.13977E-6	0.10
5.50E16	2.25E16	3.03754E-6	3.78295E-6	0.09
6.00E16	2.00E16	3.02348E-6	3.42893E-6	0.08
6.50E16	1.75E16	4.61772E-6	4.76813E-6	0.16
7.00E16	1.50E16	3.21534E-6	2.99931E-6	0.07
7.50E16	1.25E16	3.74859E-6	3.12646E-6	0.08
8.00E16	1.00E16	4.12562E-6	3.03192E-6	0.08
8.50E16	7.50E15	4.67477E-6	2.95350E-6	0.08
9.00E16	5.00E15	5.54168E-6	2.87540E-6	0.08
9.50E16	2.50E15	7.29875E-6	2.79764E-6	0.08

5.00E16	4.75E17	1.11565E-6	3.77692E-6	0.22
1.00E17	4.50E17	1.02396E-6	2.94049E-6	0.18
1.50E17	4.25E17	1.01339E-6	2.55837E-6	0.17
2.00E17	4.00E17	1.00950E-6	2.27902E-6	0.16
2.50E17	3.75E17	1.04551E-6	2.12434E-6	0.16
3.00E17	3.50E17	1.04133E-6	1.94013E-6	0.15
3.50E17	3.25E17	1.07671E-6	1.84089E-6	0.15
4.00E17	3.00E17	1.11827E-6	1.75742E-6	0.15
4.50E17	2.75E17	1.16591E-6	1.68202E-6	0.15
5.00E17	2.50E17	1.25866E-6	1.67466E-6	0.16
5.50E17	2.25E17	1.23358E-6	1.51017E-6	0.14
6.00E17	2.00E17	1.44123E-6	1.62233E-6	0.17
6.50E17	1.75E17	2.55380E-6	2.62207E-6	0.41
7.00E17	1.50E17	1.03072E-6	9.65655E-7	0.07
7.50E17	1.25E17	1.40791E-6	1.18096E-6	0.11
8.00E17	1.00E17	1.61635E-6	1.20419E-6	0.12
8.50E17	7.50E16	1.89179E-6	1.21973E-6	0.13
9.00E17	5.00E16	2.22049E-6	1.19158E-6	0.13
9.50E17	2.50E16	2.85720E-6	1.16352E-6	0.13

5.00E17	4.75E18	6.41578E-7	1.66650E-6	0.51
1.00E18	4.50E18	4.92716E-7	1.22465E-6	0.34
1.50E18	4.25E18	4.78492E-7	1.06941E-6	0.31
2.00E18	4.00E18	4.70522E-7	9.61718E-7	0.29
2.50E18	3.75E18	4.65654E-7	8.76120E-7	0.27
3.00E18	3.50E18	4.88426E-7	8.47863E-7	0.28
3.50E18	3.25E18	5.04855E-7	8.12680E-7	0.28
4.00E18	3.00E18	5.11959E-7	7.67137E-7	0.27
4.50E18	2.75E18	5.43231E-7	7.53429E-7	0.28
5.00E18	2.50E18	5.56051E-7	7.15328E-7	0.27
5.50E18	2.25E18	6.40915E-7	7.64328E-7	0.32
6.00E18	2.00E18	8.11835E-7	8.92356E-7	0.43
6.50E18	1.75E18	1.41952E-6	1.44261E-6	0.78
7.00E18	1.50E18	3.85892E-7	3.59825E-7	0.09
7.50E18	1.25E18	4.78602E-7	4.05851E-7	0.12
8.00E18	1.00E18	5.87996E-7	4.44587E-7	0.15
8.50E18	7.50E17	6.99153E-7	4.62610E-7	0.17
9.00E18	5.00E17	8.35273E-7	4.66940E-7	0.18
9.50E18	2.50E17	1.10591E-6	4.81170E-7	0.20

5.00E18	4.75E19	2.99024E-7	6.59553E-7	0.71
1.00E19	4.50E19	3.00903E-7	5.85384E-7	0.70
1.50E19	4.25E19	3.12790E-7	5.53348E-7	0.71
2.00E19	4.00E19	3.78904E-7	5.92906E-7	0.81
2.50E19	3.75E19	3.60463E-7	5.44427E-7	0.77
3.00E19	3.50E19	4.48592E-7	6.13312E-7	0.87
3.50E19	3.25E19	2.61655E-7	3.82698E-7	0.52
4.00E19	3.00E19	4.18477E-7	5.36662E-7	0.81
4.50E19	2.75E19	3.35456E-7	4.25363E-7	0.65
5.00E19	2.50E19	3.78873E-7	4.51248E-7	0.71
5.50E19	2.25E19	3.75953E-7	4.26765E-7	0.68
6.00E19	2.00E19	4.07686E-7	4.37853E-7	0.71
6.50E19	1.75E19	4.87439E-7	4.95279E-7	0.80
7.00E19	1.50E19	1.82861E-7	1.72347E-7	0.17
7.50E19	1.25E19	1.90266E-7	1.63628E-7	0.16
8.00E19	1.00E19	2.05756E-7	1.59711E-7	0.16
8.50E19	7.50E18	2.42777E-7	1.65252E-7	0.18
9.00E19	5.00E18	2.99665E-7	1.75414E-7	0.21
9.50E19	2.50E18	3.96257E-7	1.83892E-7	0.24
=======


Table 4: The same as in table 3 but for the case $Z_{1}=Z_{2}=1$.

N_1	N_2	r_{b;1}	        r_{b;2}	      \alpha_{i}
====================================================
5.00E15	9.50E16	1.95130E-6	7.13693E-6	0.04
1.00E16	9.00E16	2.47253E-6	6.48205E-6	0.06
1.50E16	8.50E16	2.53936E-6	5.50640E-6	0.06
2.00E16	8.00E16	2.61955E-6	4.89161E-6	0.06
2.50E16	7.50E16	2.69974E-6	4.43720E-6	0.06
3.00E16	7.00E16	2.53936E-6	3.75558E-6	0.05
3.50E16	6.50E16	2.88685E-6	3.83577E-6	0.06
4.00E16	6.00E16	2.99377E-6	3.60856E-6	0.06
4.50E16	5.50E16	2.84675E-6	3.12742E-6	0.05
5.00E16	5.00E16	1.87111E-6	1.87111E-6	0.02
5.50E16	4.50E16	3.12742E-6	2.84675E-6	0.05
6.00E16	4.00E16	3.60856E-6	2.99377E-6	0.06
6.50E16	3.50E16	3.83577E-6	2.88685E-6	0.06
7.00E16	3.00E16	3.75558E-6	2.53936E-6	0.05
7.50E16	2.50E16	4.43720E-6	2.69974E-6	0.06
8.00E16	2.00E16	4.89161E-6	2.61955E-6	0.06
8.50E16	1.50E16	5.50640E-6	2.53936E-6	0.06
9.00E16	1.00E16	6.48205E-6	2.47253E-6	0.06
9.50E16	5.00E15	7.13693E-6	1.95130E-6	0.04

5.00E16	9.50E17	1.05460E-6	3.38711E-6	0.11
1.00E17	9.00E17	1.02978E-6	2.57445E-6	0.10
1.50E17	8.50E17	1.06080E-6	2.21465E-6	0.10
2.00E17	8.00E17	1.09182E-6	1.97892E-6	0.10
2.50E17	7.50E17	1.12904E-6	1.81142E-6	0.10
3.00E17	7.00E17	1.16626E-6	1.68115E-6	0.10
3.50E17	6.50E17	1.14144E-6	1.49504E-6	0.09
4.00E17	6.00E17	1.24690E-6	1.48884E-6	0.10
4.50E17	5.50E17	1.48884E-6	1.62532E-6	0.13
5.00E17	5.00E17	7.32014E-7	7.32014E-7	0.03
5.50E17	4.50E17	1.62532E-6	1.48884E-6	0.13
6.00E17	4.00E17	1.48884E-6	1.24690E-6	0.10
6.50E17	3.50E17	1.49504E-6	1.14144E-6	0.09
7.00E17	3.00E17	1.68115E-6	1.16626E-6	0.10
7.50E17	2.50E17	1.81142E-6	1.12904E-6	0.10
8.00E17	2.00E17	1.97892E-6	1.09182E-6	0.10
8.50E17	1.50E17	2.21465E-6	1.06080E-6	0.10
9.00E17	1.00E17	2.57445E-6	1.02978E-6	0.10
9.50E17	5.00E16	3.38711E-6	1.05460E-6	0.11

5.00E17	9.50E18	4.49188E-7	1.31301E-6	0.18
1.00E18	9.00E18	4.31912E-7	1.01067E-6	0.16
1.50E18	8.50E18	4.29032E-7	8.55185E-7	0.15
2.00E18	8.00E18	4.40550E-7	7.71682E-7	0.15
2.50E18	7.50E18	4.52068E-7	7.11215E-7	0.15
3.00E18	7.00E18	4.66465E-7	6.62265E-7	0.15
3.50E18	6.50E18	4.83741E-7	6.24832E-7	0.15
4.00E18	6.00E18	5.01018E-7	5.93159E-7	0.15
4.50E18	5.50E18	5.18294E-7	5.64365E-7	0.15
5.00E18	5.00E18	3.08097E-7	3.08097E-7	0.05
5.50E18	4.50E18	5.64365E-7	5.18294E-7	0.15
6.00E18	4.00E18	5.93159E-7	5.01018E-7	0.15
6.50E18	3.50E18	6.24832E-7	4.83741E-7	0.15
7.00E18	3.00E18	6.62265E-7	4.66465E-7	0.15
7.50E18	2.50E18	7.11215E-7	4.52068E-7	0.15
8.00E18	2.00E18	7.71682E-7	4.40550E-7	0.15
8.50E18	1.50E18	8.55185E-7	4.29032E-7	0.15
9.00E18	1.00E18	1.01067E-6	4.31912E-7	0.16
9.50E18	5.00E17	1.31301E-6	4.49188E-7	0.18

5.00E18	9.50E19	1.89784E-7	5.02526E-7	0.27
1.00E19	9.00E19	1.81765E-7	3.95605E-7	0.24
1.50E19	8.50E19	1.77755E-7	3.35463E-7	0.22
2.00E19	8.00E19	1.77755E-7	2.98041E-7	0.21
2.50E19	7.50E19	1.87111E-7	2.83339E-7	0.22
3.00E19	7.00E19	1.92457E-7	2.65964E-7	0.22
3.50E19	6.50E19	1.93793E-7	2.45917E-7	0.21
4.00E19	6.00E19	1.95130E-7	2.28542E-7	0.20
4.50E19	5.50E19	2.12504E-7	2.29879E-7	0.22
5.00E19	5.00E19	1.38996E-7	1.38996E-7	0.09
5.50E19	4.50E19	2.29879E-7	2.12504E-7	0.22
6.00E19	4.00E19	2.28542E-7	1.95130E-7	0.20
6.50E19	3.50E19	2.45917E-7	1.93793E-7	0.21
7.00E19	3.00E19	2.65964E-7	1.92457E-7	0.22
7.50E19	2.50E19	2.83339E-7	1.87111E-7	0.22
8.00E19	2.00E19	2.98041E-7	1.77755E-7	0.21
8.50E19	1.50E19	3.35463E-7	1.77755E-7	0.22
9.00E19	1.00E19	3.95605E-7	1.81765E-7	0.24
9.50E19	5.00E18	5.02526E-7	1.89784E-7	0.27
==========


Table 5: The potential energies $U^{(1)}$ and $U^{(2)}$
in the region of electron densities $10^{16} cm^{-3} \le N_e \le 10^{20} cm^{-3}
for the case $Z_{1}=1$ and $Z_{2}=2$ at $T=3 \cdot 10^{4}$ K $; N_{1,2} in [cm^{-3}]
			
N_1 	N_2 	U^{(1)}[eV] U^{(2)}[eV]
======================================
5.00E15	4.75E16	-1.3291	-1.4490
1.00E16	4.50E16	-1.0875	-1.1936
1.50E16	4.25E16	-3.1083	-3.1559
2.00E16	4.00E16	-2.7904	-2.8365
2.50E16	3.75E16	-2.5595	-2.6040
3.00E16	3.50E16	-2.3854	-2.4277
3.50E16	3.25E16	-2.2394	-2.2860
4.00E16	3.00E16	-2.1396	-2.1735
4.50E16	2.75E16	-2.0416	-2.0702
5.00E16	2.50E16	-2.0699	-2.0942
5.50E16	2.25E16	-1.8881	-1.9077
6.00E16	2.00E16	-1.7087	-1.7238
6.50E16	1.75E16	-2.4091	-2.4127
7.00E16	1.50E16	-1.6026	-1.6102
7.50E16	1.25E16	-1.8770	-1.8820
8.00E16	1.00E16	-2.0739	-2.0740
8.50E16	7.50E15	-2.3646	-2.3574
9.00E16	5.00E15	-2.8349	-2.7820
9.50E16	2.50E15	-1.1173	-1.0953

5.00E16	4.75E17	-1.6940	-2.0187
1.00E17	4.50E17	-1.3479	-4.7729
1.50E17	4.25E17	-3.9253	-4.0906
2.00E17	4.00E17	-3.4472	-3.6157
2.50E17	3.75E17	-3.1945	-3.3586
3.00E17	3.50E17	-2.8907	-2.9999
3.50E17	3.25E17	-2.7577	-2.8415
4.00E17	3.00E17	-2.6393	-2.7089
4.50E17	2.75E17	-2.5315	-2.5889
5.00E17	2.50E17	-2.5304	-2.5764
5.50E17	2.25E17	-2.2766	-2.3161
6.00E17	2.00E17	-2.4622	-2.4913
6.50E17	1.75E17	-4.1754	-4.2149
7.00E17	1.50E17	-3.3188	-3.2513
7.50E17	1.25E17	-2.1331	-2.1319
8.00E17	1.00E17	-2.4662	-2.4446
8.50E17	7.50E16	-2.9181	-2.8643
9.00E17	5.00E16	-3.4794	-3.3616
9.50E17	2.50E16	-1.3857	-4.5200

5.00E17	4.75E18	-2.2992	-2.5614
1.00E18	4.50E18	-5.5181	-5.5583
1.50E18	4.25E18	-4.6104	-4.7543
2.00E18	4.00E18	-4.0375	-4.2491
2.50E18	3.75E18	-3.6117	-3.8726
3.00E18	3.50E18	-3.4885	-3.7548
3.50E18	3.25E18	-3.4251	-3.6082
4.00E18	3.00E18	-3.2483	-3.4173
4.50E18	2.75E18	-3.2200	-3.3600
5.00E18	2.50E18	-3.0686	-3.1981
5.50E18	2.25E18	-3.3250	-3.4108
6.00E18	2.00E18	-3.9886	-3.9973
6.50E18	1.75E18	-2.1292	-7.2375
7.00E18	1.50E18	-3.4402	-3.3824
7.50E18	1.25E18	-4.5265	-4.4823
8.00E18	1.00E18	-2.5547	-6.0326
8.50E18	7.50E17	-3.0805	-2.9556
9.00E18	5.00E17	-3.7710	-3.3700
9.50E18	2.50E17	-5.3155	-4.8040

5.00E18	4.75E19	-1.9051	-7.6028
1.00E19	4.50E19	-6.8949	-6.1569
1.50E19	4.25E19	-6.2528	-5.6731
2.00E19	4.00E19	-7.3580	-6.4351
2.50E19	3.75E19	-6.3557	-5.6541
3.00E19	3.50E19	-2.1099	-7.0146
3.50E19	3.25E19	-3.7663	-3.8298
4.00E19	3.00E19	-6.4577	-5.7475
4.50E19	2.75E19	-4.5282	-4.3075
5.00E19	2.50E19	-4.9973	-4.6592
5.50E19	2.25E19	-4.6602	-4.4269
6.00E19	2.00E19	-4.8825	-4.6121
6.50E19	1.75E19	-5.9504	-5.4742
7.00E19	1.50E19	-3.9093	-3.0972
7.50E19	1.25E19	-4.2198	-3.3306
8.00E19	1.00E19	-4.8559	-3.8560
8.50E19	7.50E18	-2.5515	-5.6651
9.00E19	5.00E18	-3.1983	-3.0571
9.50E19	2.50E18	-4.5490	-5.0868
============


Table 6: The same as in table 5, but for the case $Z_{1}=Z_{2}=1$.

N_1 	N_2 	U^{(1)}[eV] U^{(2)}[eV]
======================================
5.00E15	9.50E16	-0.5272	-0.5464
1.00E16	9.00E16	-0.4828	-0.4972
1.50E16	8.50E16	-1.3967	-1.4091
2.00E16	8.00E16	-1.2308	-1.2418
2.50E16	7.50E16	-1.1140	-1.1205
3.00E16	7.00E16	-0.9375	-0.9418
3.50E16	6.50E16	-0.9597	-0.9625
4.00E16	6.00E16	-0.9018	-0.9035
4.50E16	5.50E16	-0.7791	-0.7798
5.00E16	5.00E16	-0.9807	-0.9807
5.50E16	4.50E16	-0.7798	-0.7791
6.00E16	4.00E16	-0.9035	-0.9018
6.50E16	3.50E16	-0.9625	-0.9597
7.00E16	3.00E16	-0.9418	-0.9375
7.50E16	2.50E16	-1.1205	-1.1140
8.00E16	2.00E16	-1.2418	-1.2308
8.50E16	1.50E16	-1.4091	-1.3967
9.00E16	1.00E16	-0.4972	-0.4828
9.50E16	5.00E15	-0.5464	-0.5272

5.00E16	9.50E17	-0.7593	-0.8098
1.00E17	9.00E17	-2.0181	-2.0602
1.50E17	8.50E17	-1.6999	-1.7374
2.00E17	8.00E17	-1.5007	-1.5352
2.50E17	7.50E17	-1.3765	-1.3954
3.00E17	7.00E17	-1.2759	-1.2886
3.50E17	6.50E17	-1.1304	-1.1387
4.00E17	6.00E17	-1.1285	-1.1333
4.50E17	5.50E17	-1.2401	-1.2422
5.00E17	5.00E17	-1.1436	-1.1436
5.50E17	4.50E17	-1.2422	-1.2401
6.00E17	4.00E17	-1.1333	-1.1285
6.50E17	3.50E17	-1.1387	-1.1304
7.00E17	3.00E17	-1.2886	-1.2759
7.50E17	2.50E17	-1.3954	-1.3765
8.00E17	2.00E17	-1.5352	-1.5007
8.50E17	1.50E17	-1.7374	-1.6999
9.00E17	1.00E17	-2.0602	-2.0181
9.50E17	5.00E16	-0.8098	-0.7593

5.00E17	9.50E18	-0.8753	-0.9971
1.00E18	9.00E18	-2.2750	-2.3856
1.50E18	8.50E18	-1.8441	-1.9489
2.00E18	8.00E18	-1.6310	-1.7308
2.50E18	7.50E18	-1.5274	-1.5789
3.00E18	7.00E18	-1.4250	-1.4590
3.50E18	6.50E18	-1.3474	-1.3691
4.00E18	6.00E18	-1.2811	-1.2938
4.50E18	5.50E18	-1.2200	-1.2259
5.00E18	5.00E18	-1.3367	-1.3367
5.50E18	4.50E18	-1.2259	-1.2200
6.00E18	4.00E18	-1.2938	-1.2811
6.50E18	3.50E18	-1.3691	-1.3474
7.00E18	3.00E18	-1.4590	-1.4250
7.50E18	2.50E18	-1.5789	-1.5274
8.00E18	2.00E18	-1.7308	-1.6310
8.50E18	1.50E18	-1.9489	-1.8441
9.00E18	1.00E18	-2.3856	-2.2750
9.50E18	5.00E17	-0.9971	-0.8753

5.00E18	9.50E19	-3.0124	-3.2754
1.00E19	9.00E19	-2.0239	-2.2875
1.50E19	8.50E19	-1.5810	-1.8474
2.00E19	8.00E19	-1.4308	-1.6055
2.50E19	7.50E19	-1.4014	-1.5153
3.00E19	7.00E19	-1.3374	-1.4127
3.50E19	6.50E19	-1.2500	-1.2989
4.00E19	6.00E19	-2.9730	-3.0020
4.50E19	5.50E19	-3.0185	-3.0306
5.00E19	5.00E19	-1.3567	-1.3567
5.50E19	4.50E19	-3.0306	-3.0185
6.00E19	4.00E19	-3.0020	-2.9730
6.50E19	3.50E19	-1.2989	-1.2500
7.00E19	3.00E19	-1.4127	-1.3374
7.50E19	2.50E19	-1.5153	-1.4014
8.00E19	2.00E19	-1.6055	-1.4308
8.50E19	1.50E19	-1.8474	-1.5810
9.00E19	1.00E19	-2.2875	-2.0239
9.50E19	5.00E18	-3.2754	-3.0124
==================

Table 7: The same as in table 5, but only for
N_e= 10^{19} cm^{-3}
and for the temperature $T=1 \cdot 10^{4}$ K $.

N_1 		N_2 	U^{(1)}[eV] 	U^{(2)}[eV]
==================================================
5.00E+17	4.75E+18	-1.9773	-1.5308
1.00E+18	4.50E+18	-1.6876	-1.3349
1.50E+18	4.25E+18	-1.3688	-1.1507
2.00E+18	4.00E+18	-1.3338	-1.1100
2.50E+18	3.75E+18	-1.1501	-1.0134
3.00E+18	3.50E+18	-1.2031	-1.0356
3.50E+18	3.25E+18	-1.3549	-1.1197
4.00E+18	3.00E+18	-1.1948	-1.0428
4.50E+18	2.75E+18	-1.1384	-1.2928
5.00E+18	2.50E+18	-1.3702	-1.1655
5.50E+18	2.25E+18	-1.0453	-1.2148
6.00E+18	2.00E+18	-1.0784	-1.2665
6.50E+18	1.75E+18	-1.6423	-1.3926
7.00E+18	1.50E+18	-1.0080	-0.9947
7.50E+18	1.25E+18	-1.2010	-1.2802
8.00E+18	1.00E+18	-1.5395	-1.8467
8.50E+18	7.50E+17	-2.0302	-1.3173
9.00E+18	5.00E+17	-0.8279	-1.9363
9.50E+18	2.50E+17	-1.2046	-0.9980
=======

Table 8: The same as in table 6, but only for N_e=10^{19} cm^{-3}
and for the case  $Z_{1}=Z_{2}=1$ and temperature $T=1 \cdot 10^{4}$ K $.

N_1 		N_2 	U^{(1)}[eV] 	U^{(2)}[eV]
==================================================
5.00E+17	9.50E+18	-0.8208	-0.9394
1.00E+18	9.00E+18	-0.5228	-0.6459
1.50E+18	8.50E+18	-0.3943	-0.5211
2.00E+18	8.00E+18	-0.3894	-0.4661
2.50E+18	7.50E+18	-0.3684	-0.4189
3.00E+18	7.00E+18	-0.3492	-0.3820
3.50E+18	6.50E+18	-1.0340	-1.0545
4.00E+18	6.00E+18	-0.9412	-0.9532
4.50E+18	5.50E+18	-0.9471	-0.9523
5.00E+18	5.00E+18	-0.3400	-0.3400
5.50E+18	4.50E+18	-0.9523	-0.9471
6.00E+18	4.00E+18	-0.9532	-0.9412
6.50E+18	3.50E+18	-1.0545	-1.0340
7.00E+18	3.00E+18	-0.3820	-0.3492
7.50E+18	2.50E+18	-0.4189	-0.3684
8.00E+18	2.00E+18	-0.4661	-0.3894
8.50E+18	1.50E+18	-0.5211	-0.3943
9.00E+18	1.00E+18	-0.6459	-0.5228
9.50E+18	5.00E+17	-0.9394	-0.8208
====

Table 9: The same as in table 7, but for the temperature
$T=1.5 \cdot 10^{4}$ K $.

N_1 		N_2 	U^{(1)}[eV] 	U^{(2)}[eV]
==================================================
5.00E+17	4.75E+18	-0.9980	-3.8962
1.00E+18	4.50E+18	-0.9730	-3.5146
1.50E+18	4.25E+18	-3.1747	-2.9271
2.00E+18	4.00E+18	-1.0502	-3.5926
2.50E+18	3.75E+18	-1.1136	-3.9718
3.00E+18	3.50E+18	-1.1271	-3.9895
3.50E+18	3.25E+18	-1.9135	-1.9608
4.00E+18	3.00E+18	-1.1644	-4.1605
4.50E+18	2.75E+18	-2.3361	-2.2399
5.00E+18	2.50E+18	-2.5172	-2.3754
5.50E+18	2.25E+18	-2.4370	-2.3209
6.00E+18	2.00E+18	-2.3205	-2.2406
6.50E+18	1.75E+18	-2.3875	-2.3035
7.00E+18	1.50E+18	-1.8557	-1.5105
7.50E+18	1.25E+18	-2.0927	-1.7081
8.00E+18	1.00E+18	-2.5541	-2.1161
8.50E+18	7.50E+17	-1.3182	-3.0135
9.00E+18	5.00E+17	-1.6232	-1.5941
9.50E+18	2.50E+17	-2.3630	-2.7352
======

Table 10: The same as in table 8, but for the case  $Z_{1}=Z_{2}=1$
and temperature $T=1.5 \cdot 10^{4}$ K $.

N_1 		N_2 	U^{(1)}[eV] 	U^{(2)}[eV]
==================================================
5.00E+17	9.50E+18	-1.4973	-1.6233
1.00E+18	9.00E+18	-1.0381	-1.1612
1.50E+18	8.50E+18	-0.8401	-0.9607
2.00E+18	8.00E+18	-0.7577	-0.8385
2.50E+18	7.50E+18	-0.7357	-0.7884
3.00E+18	7.00E+18	-0.6615	-0.6982
3.50E+18	6.50E+18	-0.6515	-0.6746
4.00E+18	6.00E+18	-0.6421	-0.6548
4.50E+18	5.50E+18	-1.5021	-1.5076
5.00E+18	5.00E+18	-0.6630	-0.6630
5.50E+18	4.50E+18	-1.5076	-1.5021
6.00E+18	4.00E+18	-0.6548	-0.6421
6.50E+18	3.50E+18	-0.6746	-0.6515
7.00E+18	3.00E+18	-0.6982	-0.6615
7.50E+18	2.50E+18	-0.7884	-0.7357
8.00E+18	2.00E+18	-0.8385	-0.7577
8.50E+18	1.50E+18	-0.9607	-0.8401
9.00E+18	1.00E+18	-1.1612	-1.0381
9.50E+18	5.00E+17	-1.6233	-1.4973
=====

Table 11: The same as in table 7, but for the
temperature $T=2  \cdot 10^{4}$ K $.

N_1 		N_2 	U^{(1)}[eV] 	U^{(2)}[eV]
==================================================
5.00E+17	4.75E+18	-1.5298	-1.7147
1.00E+18	4.50E+18	-3.9406	-3.8593
1.50E+18	4.25E+18	-3.3177	-3.3599
2.00E+18	4.00E+18	-1.9709	-1.9472
2.50E+18	3.75E+18	-2.4853	-2.7179
3.00E+18	3.50E+18	-2.4365	-2.6135
3.50E+18	3.25E+18	-2.3361	-2.5031
4.00E+18	3.00E+18	-2.3699	-2.5016
4.50E+18	2.75E+18	-2.3551	-2.4679
5.00E+18	2.50E+18	-2.5670	-2.6234
5.50E+18	2.25E+18	-4.0236	-3.8298
6.00E+18	2.00E+18	-4.3209	-4.0922
6.50E+18	1.75E+18	-3.4803	-3.3774
7.00E+18	1.50E+18	-2.4317	-2.1772
7.50E+18	1.25E+18	-2.7272	-2.4382
8.00E+18	1.00E+18	-3.6079	-3.4171
8.50E+18	7.50E+17	-1.9337	-2.0996
9.00E+18	5.00E+17	-2.4097	-2.5540
9.50E+18	2.50E+17	-3.3652	-2.8270
======

Table 12: The same as in table 8, but for the case  $Z_{1}=Z_{2}=1$
and temperature $T=2  \cdot 10^{4}$ K $.

N_1 		N_2 	U^{(1)}[eV] 	U^{(2)}[eV]
==================================================
5.00E+17	9.50E+18	-2.1323	-2.2580
1.00E+18	9.00E+18	-1.4888	-1.6073
1.50E+18	8.50E+18	-1.2149	-1.3300
2.00E+18	8.00E+18	-1.0686	-1.1506
2.50E+18	7.50E+18	-0.9951	-1.0494
3.00E+18	7.00E+18	-0.9722	-1.0076
3.50E+18	6.50E+18	-0.9236	-0.9463
4.00E+18	6.00E+18	-0.8533	-0.8663
4.50E+18	5.50E+18	-0.8208	-0.8265
5.00E+18	5.00E+18	-0.9534	-0.9534
5.50E+18	4.50E+18	-0.8265	-0.8208
6.00E+18	4.00E+18	-0.8663	-0.8533
6.50E+18	3.50E+18	-0.9463	-0.9236
7.00E+18	3.00E+18	-1.0076	-0.9722
7.50E+18	2.50E+18	-1.0494	-0.9951
8.00E+18	2.00E+18	-1.1506	-1.0686
8.50E+18	1.50E+18	-1.3300	-1.2149
9.00E+18	1.00E+18	-1.6073	-1.4888
9.50E+18	5.00E+17	-2.2580	-2.1323
=====

Table 13: The same as in table 7, but for the temperature
$T=2.5  \cdot 10^{4}$ K $.

N_1 		N_2 	U^{(1)}[eV] 	U^{(2)}[eV]
==================================================
5.00E+17	4.75E+18	-1.9285	-2.1609
1.00E+18	4.50E+18	-4.7046	-4.7062
1.50E+18	4.25E+18	-3.8794	-4.0100
2.00E+18	4.00E+18	-3.4231	-3.6246
2.50E+18	3.75E+18	-3.1280	-3.3724
3.00E+18	3.50E+18	-3.0209	-3.2214
3.50E+18	3.25E+18	-2.9400	-3.1140
4.00E+18	3.00E+18	-2.9128	-3.0554
4.50E+18	2.75E+18	-2.7842	-2.9213
5.00E+18	2.50E+18	-2.8286	-2.9391
5.50E+18	2.25E+18	-2.7654	-2.8653
6.00E+18	2.00E+18	-3.7752	-3.7384
6.50E+18	1.75E+18	-1.8925	-6.4790
7.00E+18	1.50E+18	-2.9244	-2.8393
7.50E+18	1.25E+18	-3.6153	-3.5124
8.00E+18	1.00E+18	-2.1358	-4.9840
8.50E+18	7.50E+17	-2.5575	-2.8893
9.00E+18	5.00E+17	-3.1096	-2.6989
9.50E+18	2.50E+17	-4.3811	-3.8468
===

Table 14: The same as in table 8, but for the case  $Z_{1}=Z_{2}=1$
and temperature $T=2.5  \cdot 10^{4}$ K $.

N_1 		N_2 	U^{(1)}[eV] 	U^{(2)}[eV]
==================================================
5.00E+17	9.50E+18	-0.7605	-0.8746
1.00E+18	9.00E+18	-1.9144	-2.0288
1.50E+18	8.50E+18	-1.5577	-1.6671
2.00E+18	8.00E+18	-1.3813	-1.4865
2.50E+18	7.50E+18	-1.3001	-1.3519
3.00E+18	7.00E+18	-1.1809	-1.2162
3.50E+18	6.50E+18	-1.1568	-1.1793
4.00E+18	6.00E+18	-1.0647	-1.0778
4.50E+18	5.50E+18	-1.0536	-1.0592
5.00E+18	5.00E+18	-1.1760	-1.1760
5.50E+18	4.50E+18	-1.0592	-1.0536
6.00E+18	4.00E+18	-1.0778	-1.0647
6.50E+18	3.50E+18	-1.1793	-1.1568
7.00E+18	3.00E+18	-1.2162	-1.1809
7.50E+18	2.50E+18	-1.3519	-1.3001
8.00E+18	2.00E+18	-1.4865	-1.3813
8.50E+18	1.50E+18	-1.6671	-1.5577
9.00E+18	1.00E+18	-2.0288	-1.9144
9.50E+18	5.00E+17	-0.8746	-0.7605
====
\end{verbatim}

\end{document}